\documentclass[useAMS,usenatbib,usegraphicx]{mn2e}

\title[The C~II UV feature in TTSs]{Constraints to the magnetospheric properties of T Tauri stars - I.
The  C~II], Fe~II] and Si~II]  ultraviolet features}

\author[L\'opez-Mart\'inez, F. and G\'omez de Castro, A.I.]{Fatima L\'opez-Mart\'{i}nez$^1$ \thanks{E-mail: fatimalopezmar@gmail.com} and Ana In\'es G\'omez de Castro$^1$ \thanks{E-mail: aig@ucm.es}\\  
$^1$AEGORA Research Group, Universidad Complutense de Madrid, Plaza de Ciencias 3,
28040 Madrid, Spain }

\begin{document}

\date{Submission, February XXth, 2014}

\pagerange{\pageref{firstpage}--\pageref{lastpage}} \pubyear{}

\maketitle

\label{firstpage}

\begin{abstract}
The C~II] feature at $\sim 2325$~\AA\ is very prominent in the spectra
of T Tauri stars (TTSs). This feature is a quintuplet of semiforbidden
transitions excited at electron temperatures around 10,000~K that, together
with the nearby Si~II] and Fe~II] features, provides a reliable optically thin tracer 
for accurate measurement of the plasma properties
in the magnetospheres of TTSs. The spectra of 
20 (out of 27) TTSs observed with the Space Telescope Imaging Spectrograph
on board the \textit{Hubble Space Telescope} have good enough signal-to-noise-ratio at the C~II] wavelength.
For these stars, we have determined electron densities ($n_{\rm e}$) and temperatures ($T_{\rm e}$) in the line emission region 
as well as the profile broadening ($\sigma$).  
For most of the stars in the sample (17) we obtain $10^{4.1} \la T_{\rm e} \la 10^{4.5}$~K and
$10^8 \la n_{\rm e} \la 10^{12}$~cm$^{-3}$.
These stars have suprathermal line broadening ($35 \la \sigma \la 165$~km~s$^{-1}$),
except TW~Hya and CY~Tau with thermal line broadening.
Both C~II] line luminosity and broadening are found to correlate with the accretion rate.
Line emission seems to be produced in the magnetospheric accretion flow, close to the disc.
There are three exceptions: DG~Tau, RY~Tau and FU~Ori.  
The line centroids are blueshifted indicating that the line emission in these 
three stars is dominated by the outflow.
\end{abstract}

\begin{keywords}
stars: magnetic field - stars: pre-main sequence - stars: winds, outflows - ultraviolet: stars
\end{keywords}

\section{Introduction}

T~Tauri stars (TTSs) are young and low-mass ($\la 3$~M$_{\sun}$)
pre-main sequence stars with strong and complex magnetic fields and
a surrounding disc that is truncated near the corotation radius by
interaction with the magnetic field. From the observational point
of view TTSs are split into two main groups: Classical TTSs (CTTSs) and
Weak lined TTSs (WTTSs). CTTSs are accreting mass from the disc whereas
WTTSs have no or very little spectral signatures of accretion. The
material in the inner part of the disc is ionized by the stellar
radiation and channelled through the magnetic field lines
\citep{uchida1984,koenigl1991}. The gas from the disc is
accelerated to almost free-fall velocity before it reaches
the stellar surface forming an accretion shock
\citep[see, e.g., the reviews by][]{bouvier2007,aig2013a}.
Detailed simulations of the interaction between the stellar field and
the inner disc show a  complex dynamics of the magnetospheric
flow that depends on the field properties and its
stability \citep{romanova2012,kurosawa2013}.
Some analytical expressions for the hotspot shapes and the 
magnetospheric radius have been provided by \citet{kulkarni2013}.

The interaction between the star, disc and magnetic field
produces an excess emission at different wavelengths that
affects the evolution of the disc itself and the
circumstellar environment. The atmospheric and magnetospheric
energy output is released mainly in the ultraviolet (UV) spectral
range. Thus, there is a relatively large number of spectral
features in the UV that can be used as potential
tracers of the physical conditions in TTS.
Different emission lines in the UV wavelength range
provide different information about the regions in which
they are formed, the involved physical processes and the
 system geometry. For example, the Mg~II resonance
doublet at 2795.5 and 2802.7~\AA\ is produced in the chromosphere of TTS
and it is one of the strongest features in UV spectra of TTS. Mg~II is
sensitive to, and can be used as a good tracer of, atmosphere and
outflow/wind in TTS
\citep[][Lopez-Martinez \& G\'omez de Castro, submitted]{ardila2002b,calvet2004,ingleby2013}.
N~V, C~IV, He~II and Si~IV are good tracers of hot gas and
accretion processes in TTSs. The relationship between these
lines and mass accretion in TTSs has been already studied
by different authors \citep{johnskrull2000,ardila2002a,
ingleby2011,yang2012,aig2012,ardila2013,aig2013b}.

The semiforbidden lines of the C~II] quintuplet
(wavelengths: $2324.21$, $2325.4$, $2326.11$, $2327.64$, $2328.83$~\AA)
are not observed in WTTSs; however, they are readily detected
in CTTSs, even in low mass accretors \citep{lamzin2000}.
This multiplet seems to be a very sensitive tracer of accretion
or  outflows \citep{calvet2004,aig2005,ingleby2013}.
\citet{calvet2004} and \citet{ingleby2013} analysed these lines in low resolution spectra and found
a relationship between the C~II] luminosity and the accretion
luminosity.
The study of the C~II] flux ratios within a small range of
wavelengths provides a good opportunity to investigate TTS
properties because they are optically thin and their ratios
do not depend on the geometry of the accretion system and are
only slightly affected by the large uncertainties associated
with extinction determination. It is known that the relative
intensities of the emission lines of the C~II]
multiplet are sensitive to the electron density in the
range $10^8 \la n_{\rm e} \la 10^{10}$~cm$^{-3}$
\citep{stencel1981,hayes1984a,hayes1984b,keenan1986}.
Plasma in the magnetospheres and atmospheres of 
CTTSs is within this density range.  However, line blending makes
it difficult to identify the individual features and to measure the
lines ratios \citep[see,e.g.][observations of RU~Lup and DR~Tau, respectively]{lamzin2000,kravtsova2002}.

In this work, we present for the first time 
a study of C~II] line ratios in a sample of 20 CTTSs
using 30 medium-resolution spectra. We found
the best-fitting spectrum to the data using 
a grid of simulated profiles computed for a broad range  
of electron densities and temperatures.
The log of observations, the characteristics of the 
CTTSs sample and the profiles are described in Section~\ref{sample}.
The numerical method used to derive the individual
lines fluxes and the properties of the radiating plasma
is presented in Section~\ref{plasma}, that also includes 
the limitations of the method and the final results. 
In Section~4, we present the plasma properties obtained with our procedure
and they are compared with the accretion rates derived from \citet{ingleby2013}. 
To conclude, in Section~5, we provide a brief summary of the main results. 


\section{The C~II] profile of CTTSs}
\label{sample}

Our sample consists of the 27 CTTSs observed with the Space Telescope Imaging Spectrograph (STIS) on board
the \textit{Hubble Space Telescope} (\textit{HST});  no C~II] emission is detected in WTTSs.
Most of the sources (17 of 27) are located in Taurus-Auriga Molecular Cloud. The rest of the sources are in $\eta$ Chamaleon (2), $\epsilon$ Chamaleon (1), Chamaleon I (2), TW Hydra Association (2), Orion (1) and Upper Scorpius (1). DK~Tau, HN~Tau \citep{correia2006}, CV~Cha \citep{bary2008} and UX~Tau \citep{nguyen2012} are binaries with companions at distances of $2.304$, $3.109$, $11.4$ and $5.9$ arcsec, respectively, that are resolved by STIS. T~Tau \citep{furlan2006}, FU~Ori \citep{wang2004} and DF~tau \citep{unruh1998} are close binaries at distances $0.7$, $0.5$ and $0.09$ arcsec, respectively. CS~Cha is a spectroscopic binary \citep{guenther2007}. Several stars show evidence of transitional discs, but they are still accreting: CS~Cha, DM~Tau, GM~Aur, TW~Hya and UX~Tau \citep{espaillat2010}. In some sources of our sample jets/outflows have been detected: RY~Tau \citep{stonge2008}, DG~Tau \citep{coffey2008}, T~Tau \citep{furlan2006}, SZ~102 \citep{comeron2011}, AA~Tau, DF~Tau, HN~Tau and SU~Aur \citep{howard2013}.\\

\begin{figure*}
\begin{tabular}{cc}
\includegraphics[width=8cm]{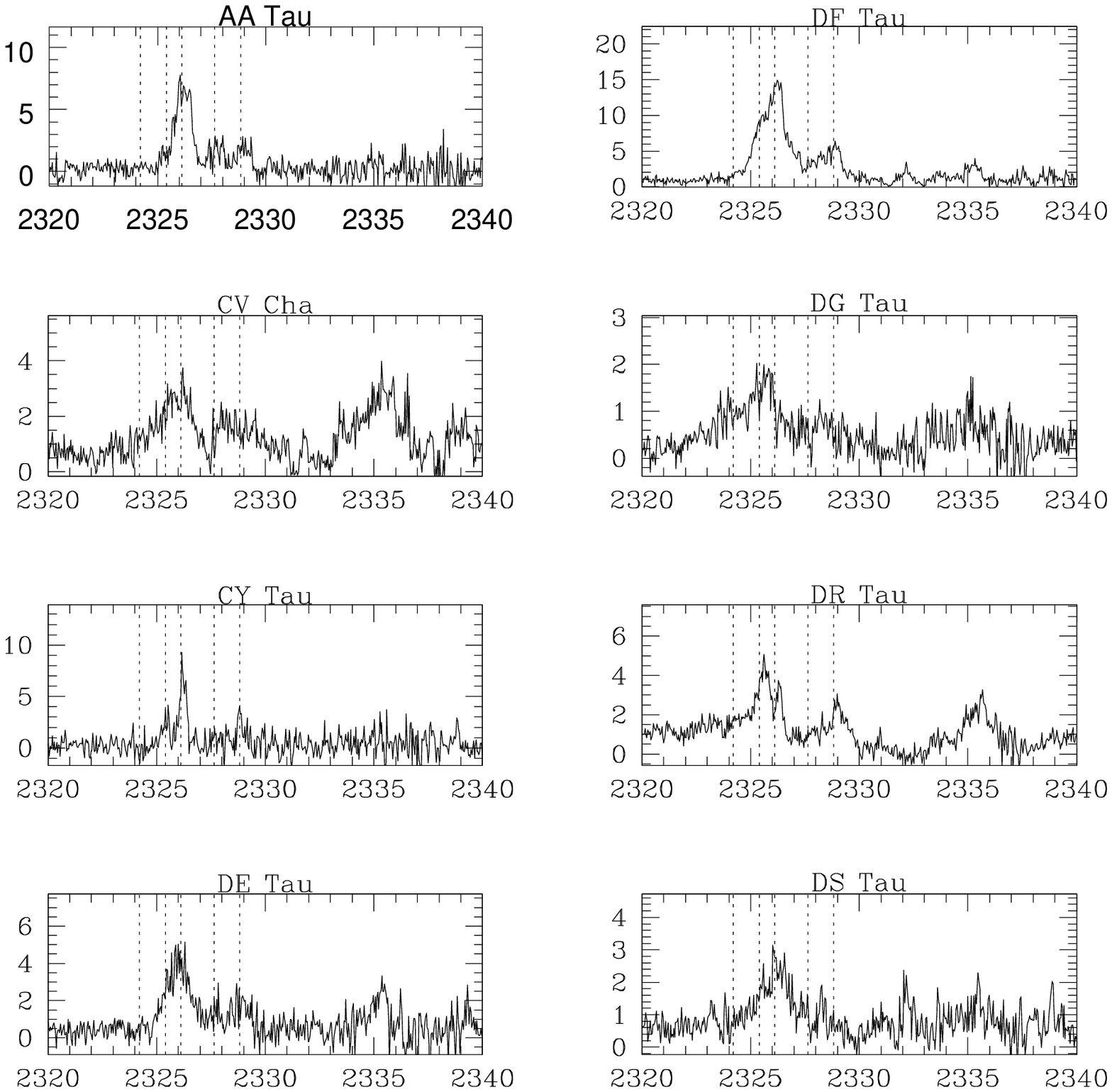} & \includegraphics[width=8cm]{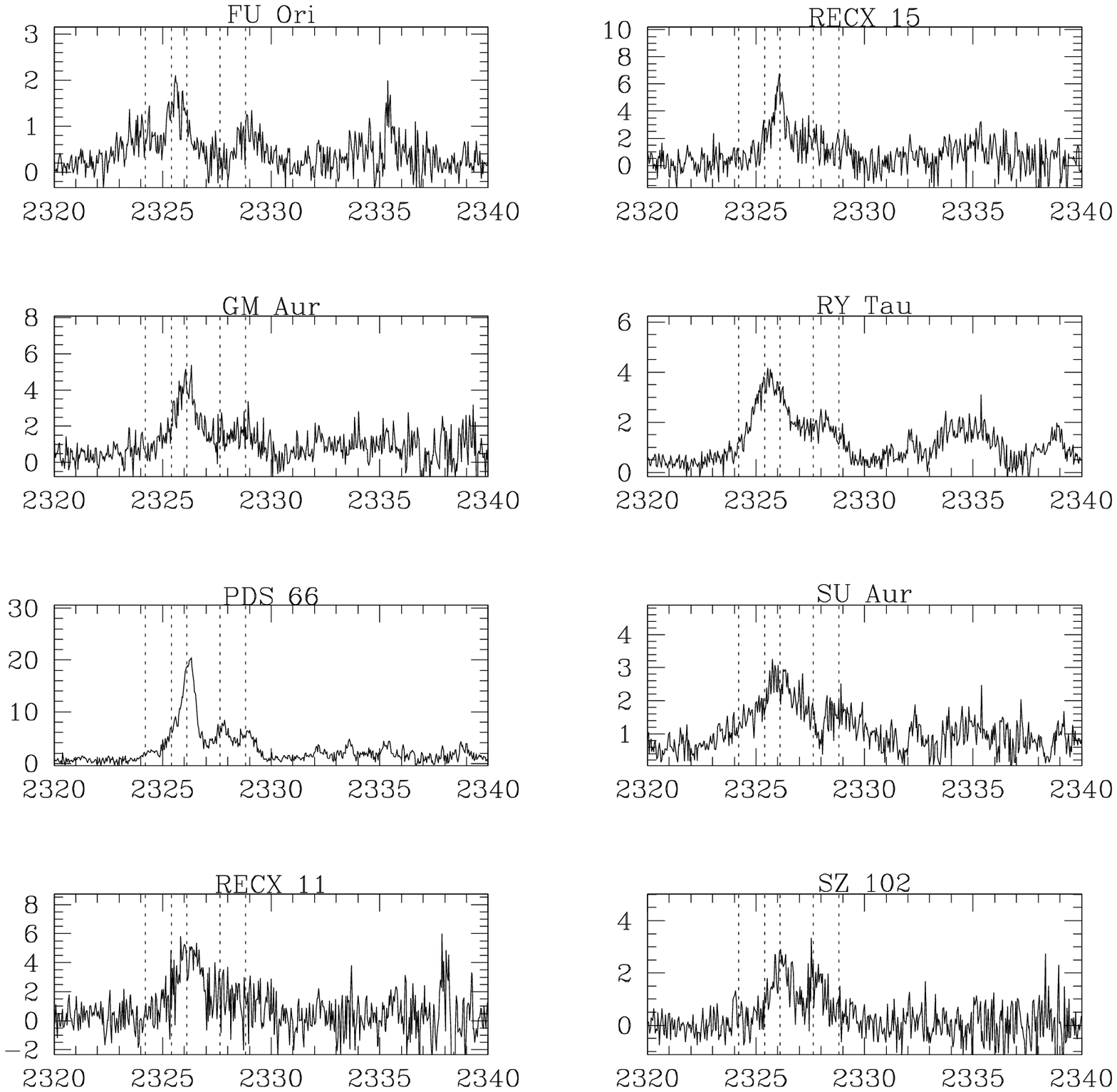} \\
\includegraphics[width=8cm]{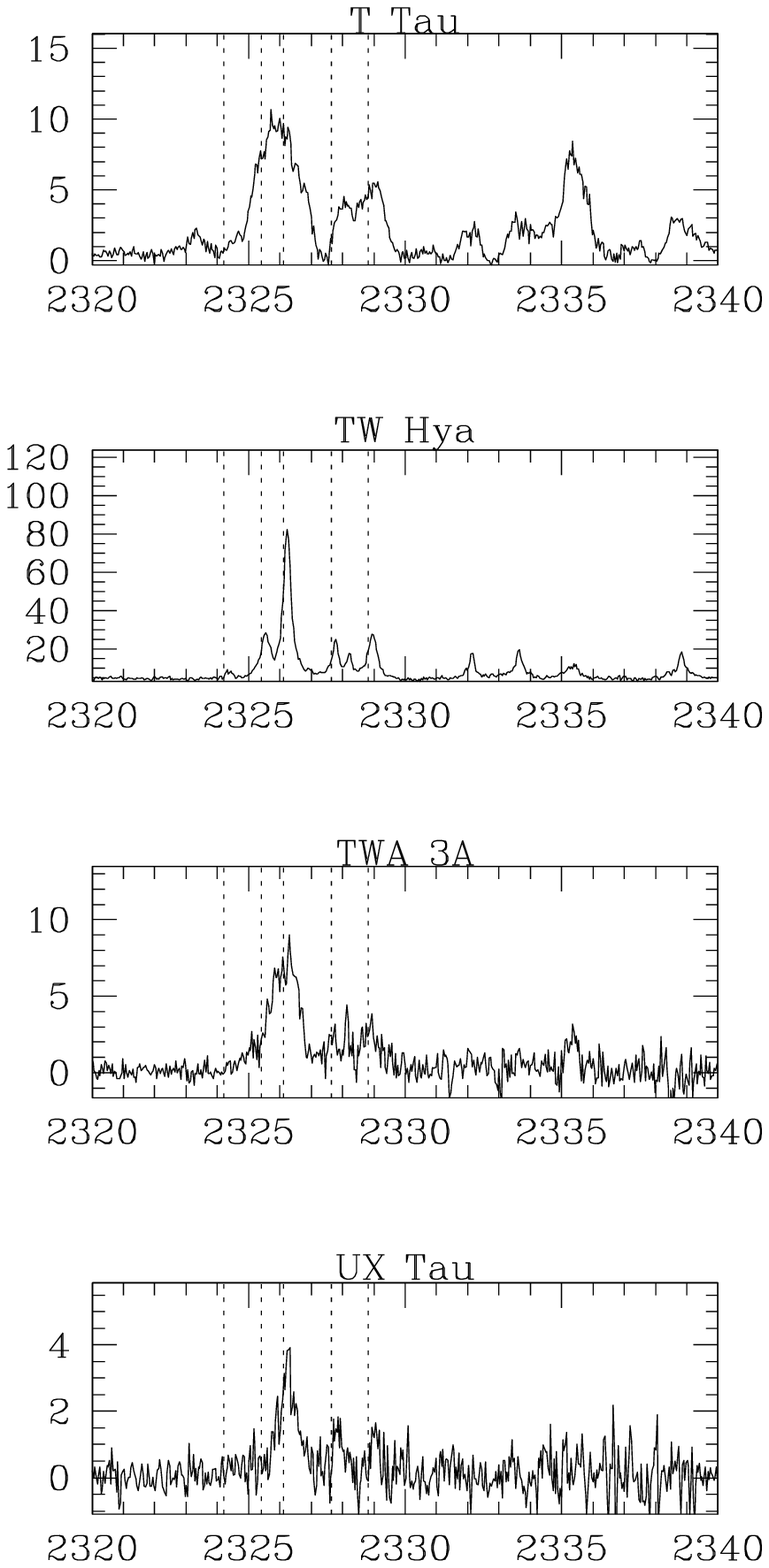} \\
\end{tabular}
\caption{The C~II] multiplet in the TTSs; only profiles with S/N $>$ 2 are plotted. The fluxes are in units of 
$10^{-14}$~erg~ s$^{-1}$~cm$^{-2}$~\AA$^{-1}$. Dashed lines mark the rest wavelengths of  the C~II] transitions. 
For stars with multiple observations, the 
spectrum with the best S/N is shown.}
\label{f1}
\end{figure*}

The sample is formed of 42 medium-resolution ($R\simeq 30 000$) spectra obtained with grating E230M; the log of data is provided in Table~\ref{tab1}.
\begin{table}
\caption{Log of observations. \label{tab1}}
\begin{tabular}{ccccc}
\hline
Star   & Obs date   & Data set & Exp time  & S/N  \\
       & (yy/mm/dd) &         & (s)		& \\ \hline
AA Tau	&	11/01/07	&	ob6ba7030	&	1462.2	&	3.21	\\	\hline
CS Cha	&	11/06/01	&	ob6bb6030	&	1785.2	&	1.65	\\	\hline
CV Cha	&	11/04/13	&	ob6b18020	&	2598.2	&	3.30	\\	\hline
CY Tau	&	00/12/06	&	o5cf03020	&	738	&	2.54	\\	
		&	00/12/06	&	o5cf03030	&	282	&	1.75	\\	\hline
DE	Tau	&	10/08/20	&	ob6ba8030	&	1388.1	&	3.58	\\	\hline
DF Tau	&	99/09/18	&	o5kc01020	&	1670.2	&	14.72	\\	\hline
DG Tau	&	01/02/20	&	o63l03010	&	2345	&	1.87	\\	
		&	01/02/20	&	o63l03020	&	2923	&	2.66	\\	
		&	01/02/20	&	o63l03030	&	2923	&	2.56	\\	
		&	01/02/20	&	o63l03040	&	2923	&	1.91	\\	\hline
DK Tau	&	10/02/04	&	ob6bb2030	&	854.4	&	0.81	\\	\hline
DM Tau	&	10/08/22	&	ob6ba2030	&	1330.1	&	1.37	\\	\hline
DN Tau	&	11/09/10	&	ob6ba4030	&	1441.2	&	1.72	\\	\hline
DR Tau	&	00/08/29	&	o5cf02020	&	916	&	1.12	\\	
		&	01/02/09	&	o63l04010	&	2327	&	2.04	\\	
		&	01/02/09	&	o63l04020	&	2880	&	2.26	\\	
		&	10/02/15	&	ob6bb4030	&	881.3	&	0.44	\\	\hline
DS Tau	&	00/08/24	&	o5cf01020	&	878	&	2.06	\\	
		&	01/02/23	&	o63l08010	&	2345	&	2.27	\\	
		&	01/02/23	&	o63l08020	&	2923	&	2.12	\\	\hline
FM Tau	&	11/09/21	&	ob6ba0030	&	1401.2	&	0.64	\\	\hline
FU Ori	&	01/02/22	&	o63l07020	&	2880	&	2.54	\\	\hline
GM Aur	&	10/08/19	&	ob6ba1030	&	1300.5	&	3.61	\\	\hline
HN Tau	&	10/02/10	&	ob6ba9030	&	807.5	&	1.24	\\	\hline
PDS	66	&	11/05/23	&	ob6b23030	&	1725.2	&	11.68	\\	\hline
RECX15		&	10/02/05	&	ob6bb7030	&	916.4	&	2.45	\\	\hline
RECX11		&	09/12/12	&	ob6bc4030	&	697.8	&	2.32	\\	\hline
RY Tau	&	01/02/19	&	o63l01010	&	2353	&	7.47	\\	
		&	01/02/20	&	o63l01020	&	2923	&	8.09	\\	
		&	01/02/20	&	o63l01030	&	2923	&	7.92	\\	\hline
SU Aur	&	01/02/24	&	o63l05010	&	2383	&	5.04	\\	
		&	01/02/24	&	o63l05020	&	2940	&	4.21	\\	
		&	11/03/25	&	ob6bb1030	&	1489.2	&	2.33	\\	\hline
SZ 102	&	11/05/29	&	ob6bb9030	&	1469.2	&	3.12	\\	\hline
T Tau	&	01/02/21	&	o63l02010	&	2331	&	12.57	\\	
		&	01/02/21	&	o63l02020	&	2880	&	13.95	\\	
		&	01/02/22	&	o63l02030	&	2880	&	13.53	\\	\hline
TW Hya	&	00/05/07	&	o59d01020	&	1675.2	&	21.23	\\	\hline
TWA	3A	&	11/03/26	&	ob6b22030	&	1107.2	&	6.70	\\	\hline
UX Tau	&	11/11/10	&	ob6b54030	&	1408.2	&	2.35	\\	\hline
V836 Tau	&	11/02/05	&	ob6ba6030	&	1396.2	&	0.64	\\	\hline

\end{tabular}
\end{table}
We have selected spectra with signal-to-noise-ratio (S/N) $>2$; the S/N has been calculated over the whole
feature as described in Section~3.3. 
The spectra are shown in Fig.~\ref{f1}. 
No significant variations are detected in the spectrum of sources with multiple observations, except for DS~Tau (see Appendix A); note that though the C~II] flux of DS~Tau drops by a factor of 2 between two observations, no significant
profile shape variations are noticeable.

\begin{figure}
\centering
\includegraphics[width=8cm]{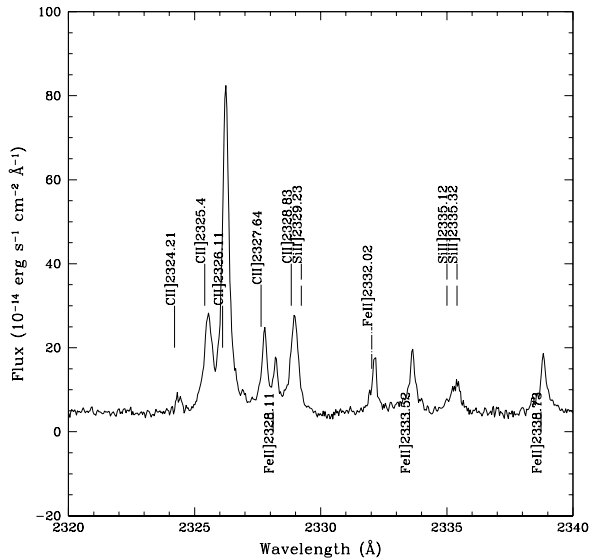}
\caption{Line identification in the spectral range 2320-2340~\AA\ on TW~Hya spectrum. \label{f2}}
\end{figure}

In Fig.\ref{f2}, the main spectral features in the 2324-2336~\AA\ range are indicated on the spectrum of TW~Hya, the star with the best S/N in the sample. Note that the C~II] multiplet is resolved.  
Additional relevant features in the range are:
\begin{enumerate}
\item  The Fe~II] lines at 2328.11 and 2333.52~\AA\ ($3d^6(\ ^5D)4s-3d^6(\ ^5D)4p$). Note that the 2328.11~\AA\ transition is blended
with the C~II] lines in most spectra. 
\item  The Fe~II] lines at 2332.02 and 2333.52~\AA . 
\item The  Si~II] multiplet at 2329.23, 2335.12 and 2335.32~\AA. 
\end{enumerate}

\section{Measuring the plasma properties}
\label{plasma}

\begin{figure}
\centering 
\includegraphics[width=8cm]{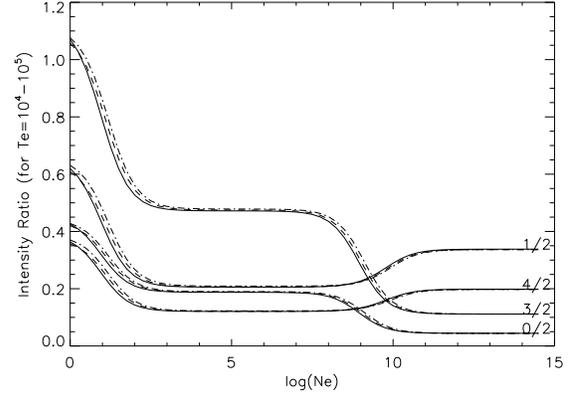}
\caption{Emissivity ratios of the C~II] lines relative to the 2326.11~\AA\ line, as a function of electron density. The labels 0,1,2,3 and 4 correspond to the C~II] lines 2324, 2325, 2326, 2327 and 2328~\AA, respectively. Solid, dashed and dotted lines correspond to temperatures of $T_{\rm e}=10^4$, $10^{4.5}$ and $T_{\rm e}=10^5$~K, respectively. \label{ratioscii}}
\end{figure}

\begin{figure}
\centering 
\includegraphics[width=8cm]{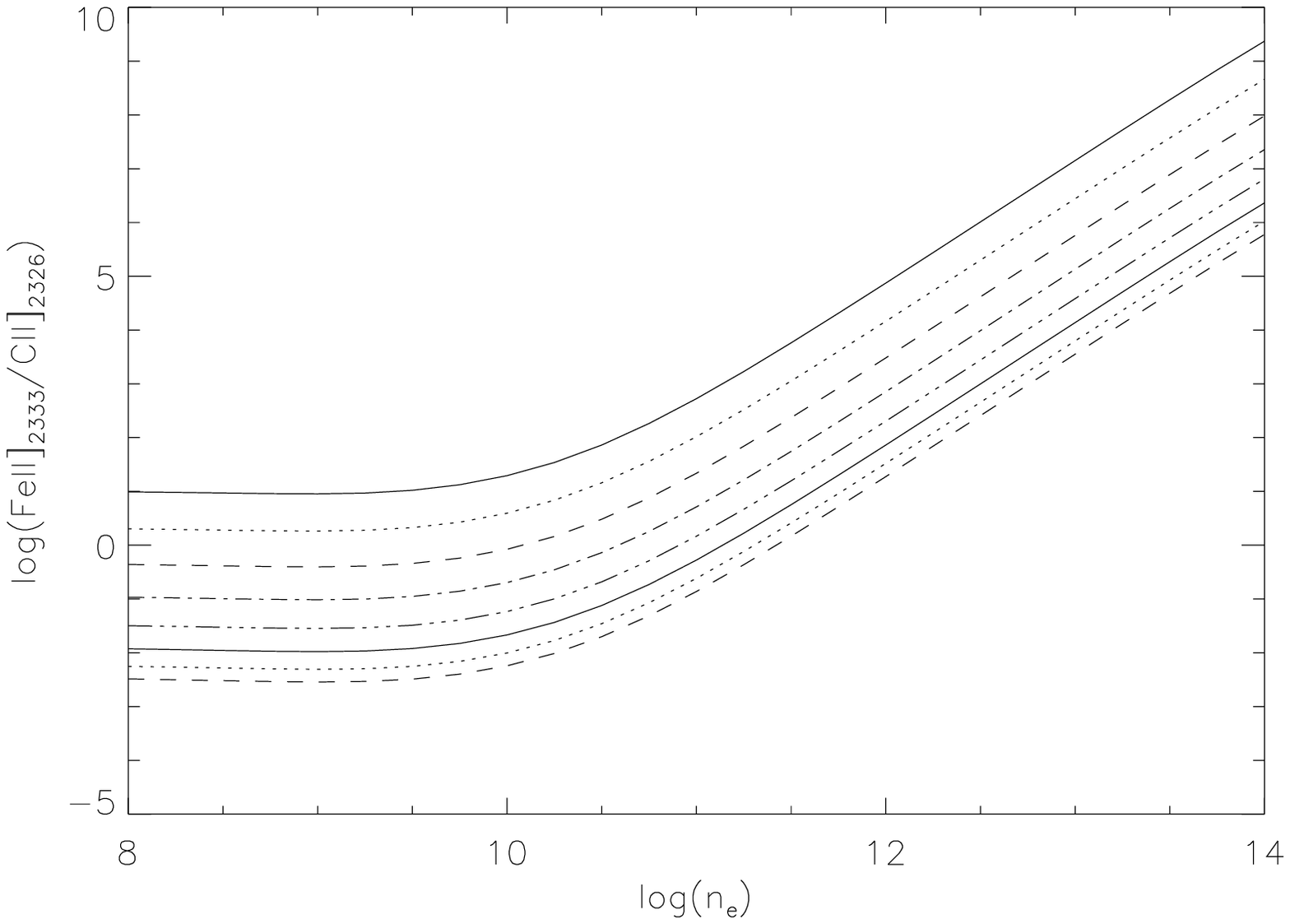}
\includegraphics[width=8cm]{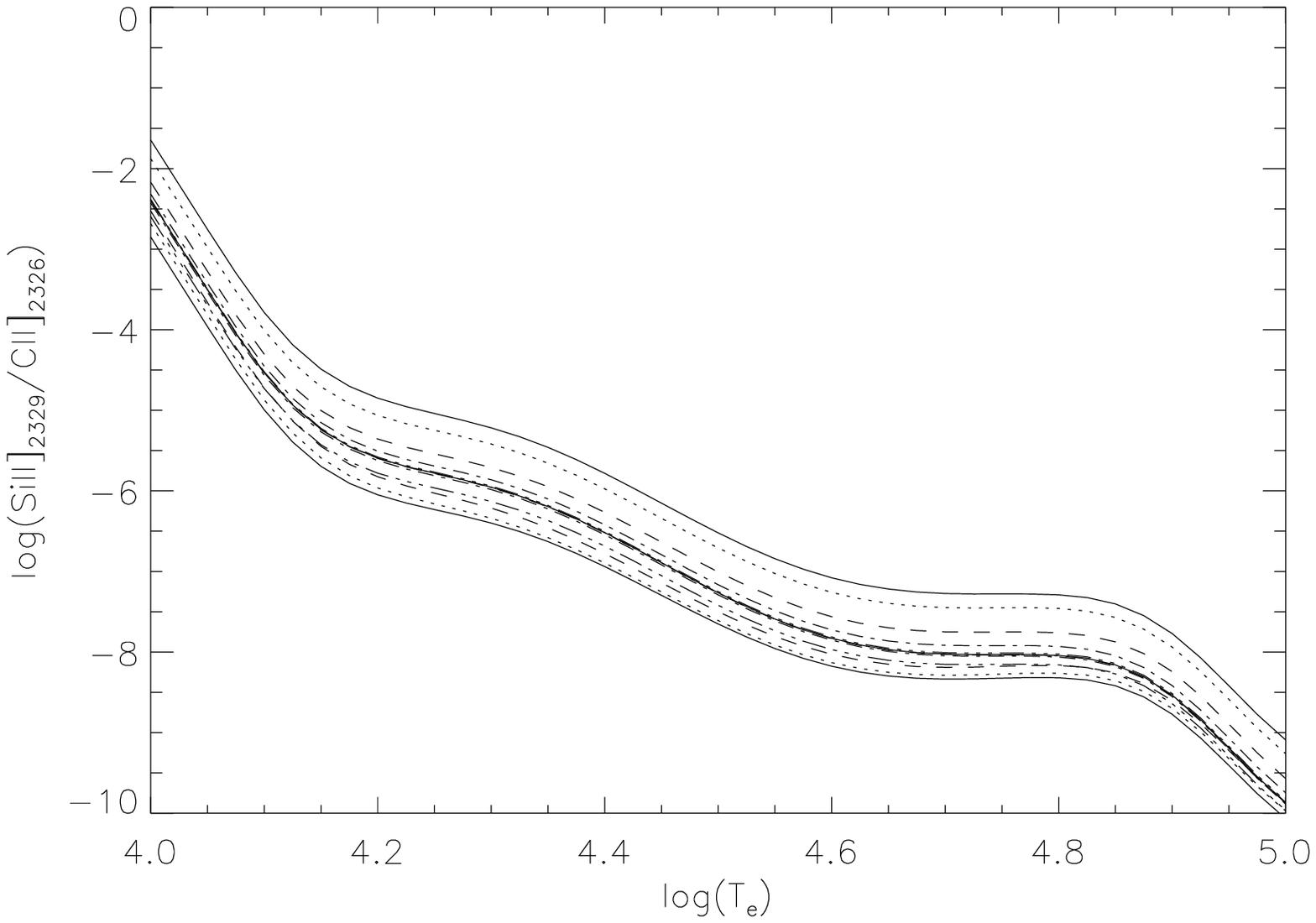}
\caption{Top panel: emissivity ratios of the Fe~II] line relative to the C~II] 2326.11~\AA\ line as a function of density for several temperatures (from $\log T_{\rm e}({\rm K})= 4.0$ to $4.175$ in steps of 0.025).
Bottom panel: emissivity ratios of the Si~II] line relative to the C~II] 2326.11~\AA\ line as a function of temperature for several densities (from $\log n_{\rm e}({\rm cm}^{-3})=0.0$ to $13.0$ in steps of 1.0). \label{ratios}}
\end{figure}

C~II], Fe~II] and  Si~II] features are intercombination transitions with very small Einstein coefficients and thus, optically thin
tracers of the radiating plasma, suitable to be used to measure directly their properties.
This characteristic was already noticed by \cite{stencel1981} for C~II] lines, who proposed to use them as electron density tracers in the 
$10^{7} \ \leq \ n_{\rm e} \ \leq 10^{10.5}$~cm$^{-3}$ range in nebulae research. In Fig.\ref{ratioscii}, we display the sensitivity of the line ratios
to $T_{\rm e}$ and $n_{\rm e}$ for this quintuplet.  
The plot was made by using the Atomic Database for Spectroscopic Diagnostics of Astrophysical Plasmas CHIANTI\footnote{www.chiantidatabase.org} \citep{dere1997,landi2013}. Note that below $n_{\rm e} \leq 10^8$~cm$~^{-3}$, the ratios are insensitive to the electron density
except for very diffuse plasmas with $n_{\rm e} \la 10^{2.5}$~cm$~^{-3}$. Therefore, other species need to be considered
to constrain the $T_{\rm e}$ of the plasma and the density for $n_{\rm e} \ga 10^{10.5}$~cm$~^{-3}$ and $n_{\rm e} \la 10^8$~cm$~^{-3}$.
The Fe~II] ratios are sensitive to the electron density for $n_{\rm e} \ga 10^9$~cm$^{-3}$ (see top panel in Fig.\ref{ratios}), the range of densities for which the C~II] quintuplet
ratios are nearly constant. 
The Si~II] ratios are more sensitive to the temperature, particularly for $T_{\rm e} \la 10^{4.5}$~K (see bottom panel in Fig.\ref{ratios}). 
The combined analysis of all these ratios yields enough information to determine unambiguously the physical properties of the region 
where the lines are formed.\\
For the calculations, we have assumed that all the lines are optically thin and formed via collisional excitation in a single plasma characterized
by a pair ($n_{\rm e}, T_{\rm e}$). CHIANTI provides the ion emissivities (erg~s$^{-1}$):
$\varepsilon_{ij} = \Delta E \,  (n_j({\rm X II})/n({\rm X II})) \, A_{ji}$, being $\Delta E$ the difference of energies between levels $j$ and $i$, $n_j({\rm X II})/n({\rm X II})$ the fraction of ions lying in the state $j$ and $A_{ji}$ the spontaneous radiative transition probability. 
The emissivities per unit volume (erg~s$^{-1}$~cm$^{-3}$) for a given ion X~II have been calculated as: 
\begin{eqnarray*}
 \epsilon_{ij} & = & \Delta E  n_j({\rm X II})  A_{ji} \nonumber \\ 
 & = & \Delta E  A_{ji}  \left(  \frac{n_j({\rm X II})}{n({\rm X II})}  \frac{n({\rm X II})}{n({\rm X})}  \frac{n({\rm X})}{n({\rm H})} \frac{n({\rm H})}{n_{\rm e}}  n_{\rm e} \right) \nonumber \\
 & = & \varepsilon_{ij} \left(\frac{n({\rm X II})}{n({\rm X})}  \frac{n({\rm X})}{n({\rm H})} \frac{n({\rm H})}{n_{\rm e}} n_{\rm e} \right),
\end{eqnarray*}
\noindent
where $n_j({\rm X II})$ is the number density of the specie X~II in the upper level ($j$), $n({\rm X II})/n({\rm X})$ is the ionization fraction of X and $n({\rm X})/n({\rm H})$ is the abundance of element X. Solar metallicity is assumed.
$n({\rm H})/n_{\rm e} = 0.83 $ has been used since $T_{\rm e} > 10^4$~K (see CHIANTI manual).

\subsection{The numerical method}

Making use of the emissivities from CHIANTI, we have computed the flux ratios relative to the C~II] (2326.11 \AA ) line of the following lines: C~II]( 2324.21, 2325.4, 2327.64 and 2328.83~\AA ), Fe II] 
(2328.11 and 2333.52~\AA), Fe~II] (2332.02~\AA) and Si~II] (2329.23, 2335.12 and 2335.32~\AA), for a grid of electron temperatures and densities. The grid covers the range 
$4.0 \leq \log T_{\rm e}({\rm K}) \leq 5.5$ and $0.0 \leq \log n_{\rm e}({\rm cm}^{-3}) \leq 14.5$ with resolutions 0.025 dex in $\log (T_{\rm e})$ and 0.25~dex in $\log (n_{\rm e})$. 
We have assumed that the lines profiles are adequately reproduced by Gaussian functions. In this manner, we have built a grid of simulated spectra in the 2323-2338~\AA\ spectral range given by

\begin{equation}
\label{eq3}
F(\lambda) =  F_0 \  \displaystyle\sum_{i=0}^{10} R_i \ \exp \left( {\frac{-(\lambda - (\lambda_i+\delta))^2}{2 \sigma^2}} \right) + F_{cont}, 
\end{equation}
\noindent where $F_0$ is the peak flux of the reference line (C~II]$_{2326}$), $R_i=F_i/F_0$ is the flux ratio between the peak of the {\it i}th line and $F_0$, $\sigma$ is the standard deviation of the Gaussian functions and $\lambda_i$ is the central wavelength of the \textit{i}th emission line (which can be shifted $\delta$~\AA\ from its expected position). $F_{cont}$ is directly computed from the observations as the average flux in the 2320-2323~\AA\ range for each spectra;
this is a featureless window (see Fig.\ref{f1}). Both dispersion ($\sigma$) and shift ($\delta$) are assumed to be the same for all lines.

\begin{figure}
\centering
\includegraphics[width=9cm]{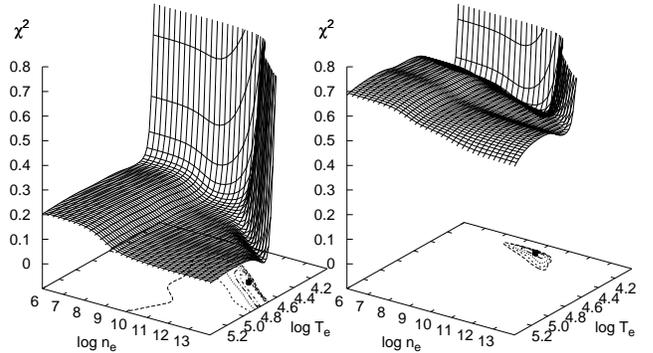}
\caption{$\chi^2$ surfaces and contours for TW~Hya (on the left) and DE~Tau (on the right). At the bottom of the figures, we projected five $\chi^2$ contours starting close to the best solution $\chi^2_{opt}$ (0.08 and 0.48, respectively), with steps of 0.01. The black point at the bottom indicates the $T_{\rm e}$ and $n_{\rm e}$ values finally adopted. \label{chi}}
\centering
\end{figure}

We developed an IDL based code to identify the synthetic spectrum that best fit the data consisting in two main steps. First, for each 
synthetic spectrum - defined by  a pair ($n_{\rm e}$, $T_{\rm e}$) - the best fit to the data is found by a least squares scheme that leaves $F_0$, $\delta$ and $\sigma$ as 
free parameters for the fit. As a result, for 
any given model $i$ ($n_{e,i}$, $T_{e,i}$), the set of parameters that best fit the data  ($F_{0,i}, \sigma _i, \delta _i$), 
as well as the residuals, $\chi^2_i$, are computed. 
This allows plotting the $\chi^2$ surface in the ($n_{\rm e}$, $T_{\rm e}$) space (see Fig.\ref{chi}). Then, the minimum of the surface 
is identified providing  the
($n_{\rm e}, T_{\rm e}$) pair that best fit the data. This minimum corresponds to the optimal fit, i.e. $\chi^2_{opt}={\rm min}(\chi^2)$.
In Table~\ref{tab3}, the $n_{\rm e}, T_{\rm e}, \sigma ,\delta$ values corresponding to the best-fitting model are provided for all the TTSs in the study. 
\begin{table*}
\caption{Physical parameters derived from the fitting. \label{tab3}}
\begin{tabular}{cccccccc}
\hline
Star  & Data set     & $\log(T_{\rm e})$   & $\log(n_{\rm e})$ & $\chi^2_{opt}$    &  $\delta$  & $\sigma$  & $F_0$ \\
       &            &       (K)     &  (cm$^{-3}$) &                &  (km s$^{-1}$)    &  (km s$^{-1}$)   & (${\rm erg}\ {\rm s}^{-1}\ {\rm cm}^{-2}$~\AA$^{-1}$)  \\
\hline
AA Tau	&	ob6ba7030	&	4.95	&	9.50	&	0.54	&	17.67	&	40.88	&	$6.70 \times 10^{-14}$		\\ \hline
CV Cha	&	ob6b18020	&	4.10	&	10.50	&	0.28	&	18.57	&	82.03	&	$1.82\times 10^{-14}$		\\ \hline
CY Tau	&	o5cf03020	&	4.50	&	11.75	&	0.49	&	15.22	&	23.73	&	$6.08\times 10^{-14}$		\\ \hline
DE Tau	&	ob6ba8030	&	4.15	&	10.00	&	0.48	&	0.90	&	56.36	&	$3.52\times 10^{-14}$	\\ \hline
DF Tau	&	o5kc0102	&	4.45	&	11.50	&	0.01	&	9.29	&	66.42	&	$1.12\times 10^{-13}$		\\ \hline
DG Tau	&	o63l03020	&	4.18	&	10.25	&	0.12	&	-67.32	&	104.72	&	$9.23\times 10^{-15}$		\\ 
		&	o63l03030	&	4.15	&	10.00	&	0.13	&	-66.03	&	116.85	&	$9.26\times 10^{-15}$		\\ \hline
DR Tau	&	o63l04010	&	5.48	&	13.75	&	0.42	&	-24.76	&	85.38	&	$1.62\times 10^{-14}$		\\ 
		&	o63l04020	&	5.48	&	13.75	&	0.48	&	-26.70	&	71.97	&	$2.02\times 10^{-14}$		\\ \hline
DS Tau	&	o5cf01020	&	4.35	&	9.75	&	0.75	&	27.34	&	62.03	&	$3.72\times 10^{-14}$		\\ 
		&	o63l08010	&	4.18	&	9.50	&	0.20	&	19.09	&	62.29	&	$1.68\times 10^{-14}$		\\ 
		&	o63l08020	&	4.18	&	9.50	&	0.15	&	16.77	&	66.03	&	$1.74\times 10^{-14}$		\\ \hline
FU Ori	&	o63l07020	&	4.18	&	10.25	&	0.14	&	-45.53	&	93.89	&	$9.12\times 10^{-15}$		\\ \hline
GM Aur	&	ob6ba1030	&	4.38	&	11.50	&	0.47	&	14.83	&	68.48	&	$3.26\times 10^{-14}$		\\ \hline
PDS66	&	ob6b23030	&	4.30	&	8.75	&	0.01	&	15.35	&	44.88	&	$1.79\times 10^{-13}$		\\ \hline
RECX15	&	ob6bb7030	&	4.15	&	8.50	&	0.86	&	1.42	&	53.91	&	$4.28\times 10^{-14}$		\\ \hline
RECX11	&	ob6bc4030	&	4.40	&	9.00	&	1.53	&	30.82	&	61.00	&	$4.31\times 10^{-14}$		\\ \hline
RY Tau	&	o63l01010	&	4.13	&	8.50	&	0.21	&	-39.34	&	95.95	&	$2.64\times 10^{-14}$		\\ 
		&	o63l01020	&	4.18	&	10.75	&	0.16	&	-30.57	&	102.14	&	$2.59\times 10^{-14}$		\\ 
		&	o63l01030	&	4.23	&	11.00	&	0.17	&	-25.67	&	110.79	&	$2.37\times 10^{-14}$		\\ \hline
SU Aur	&	o63l05010	&	4.33	&	11.00	&	0.18	&	18.70	&	158.12	&	$1.30\times 10^{-14}$		\\ 
		&	o63l05020	&	4.18	&	10.25	&	0.20	&	-6.84	&	155.80	&	$1.38\times 10^{-14}$		\\ 
		&	ob6bb1030	&	4.18	&	10.50	&	0.55	&	26.44	&	122.91	&	$1.52\times 10^{-14}$		\\ \hline
SZ102	&	ob6bb9030	&	4.45	&	1.25	&	0.35	&	25.15	&	51.72	&	$2.25\times 10^{-14}$		\\ \hline
T Tau	&	o63l02010	&	4.15	&	10.50	&	0.01	&	3.74	&	61.52	&	$9.15\times 10^{-14}$		\\ 
		&	o63l02020	&	4.13	&	10.25	&	0.01	&	3.87	&	62.16	&	$8.87\times 10^{-14}$		\\ 
		&	o63l02030	&	4.13	&	10.25	&	0.01	&	4.77	&	59.71	&	$9.18\times 10^{-14}$		\\ \hline
TWHya	&	o59d01020	&	4.50	&	12.25	&	0.07	&	15.99	&	20.25	&	$7.24\times 10^{-13}$		\\ \hline
TWA3A	&	ob6b22030	&	4.28	&	9.25	&	0.62	&	17.28	&	49.52	&	$7.19\times 10^{-13}$		\\ \hline
UX Tau	&	ob6b54030	&	4.40	&	8.75	&	0.32	&	23.34	&	35.72	&	$3.10\times 10^{-14}$		\\ \hline
\end{tabular}
\end{table*}
Initial conditions for the free parameters are set as follows: $\sigma_0 = 0.1$~\AA\ (approximately equivalent to the combination of the spectral resolution obtained with STIS/E230M and thermal broadening), $F_0$ is set as the peak flux around 2326~\AA\ and $\delta_0$ is such that $F(2326.11-\delta_0)=F_0$ in the observed spectrum. We performed several tests to check the dependence of the results on the initial values of the free parameters. By varying these initial values, the final solution ($\chi_{opt} ^2$) never differed by more than one step in the grid of $T_{\rm e}$ and $n_{\rm e}$ values. This means that the steps of the grid represent the internal precision
of the fitting procedure ($\delta \log T_{\rm e}({\rm K}) \simeq 0.025$ and $\delta \log n_{\rm e}({\rm cm}^{-3}) \simeq 0.25$); they are the same for all stars in the sample.

From the fitting procedure, we also estimated the uncertainties associated to $\delta$, $\sigma$ and each line flux. For this, 
we selected the eight closest grid points to the best fit (the local minimum)
and we calculated the standard deviation from the average value using these eight points.
The standard deviations in $\delta$ is always $\la 5$~km~s$^{-1}$, whereas in
$\sigma$ is $\la 6$~km~s$^{-1}$. These uncertainties are not provided in Table~\ref{tab3} because they are negligible.
The final simulated fluxes with their associated errors are shown in Table~\ref{tab4}.

The Fe~II]$_{2332.02}$ line has not been  considered for the fit. We have found a large discrepancy between CHIANTI predictions for the
line strength ($\epsilon(2332.02) \sim 0.06 \cdot \epsilon(2333.52)$) and the observations, where both Fe~II lines have comparable strengths (see Fig.\ref{f2}).

\begin{figure}
\includegraphics[width=8cm]{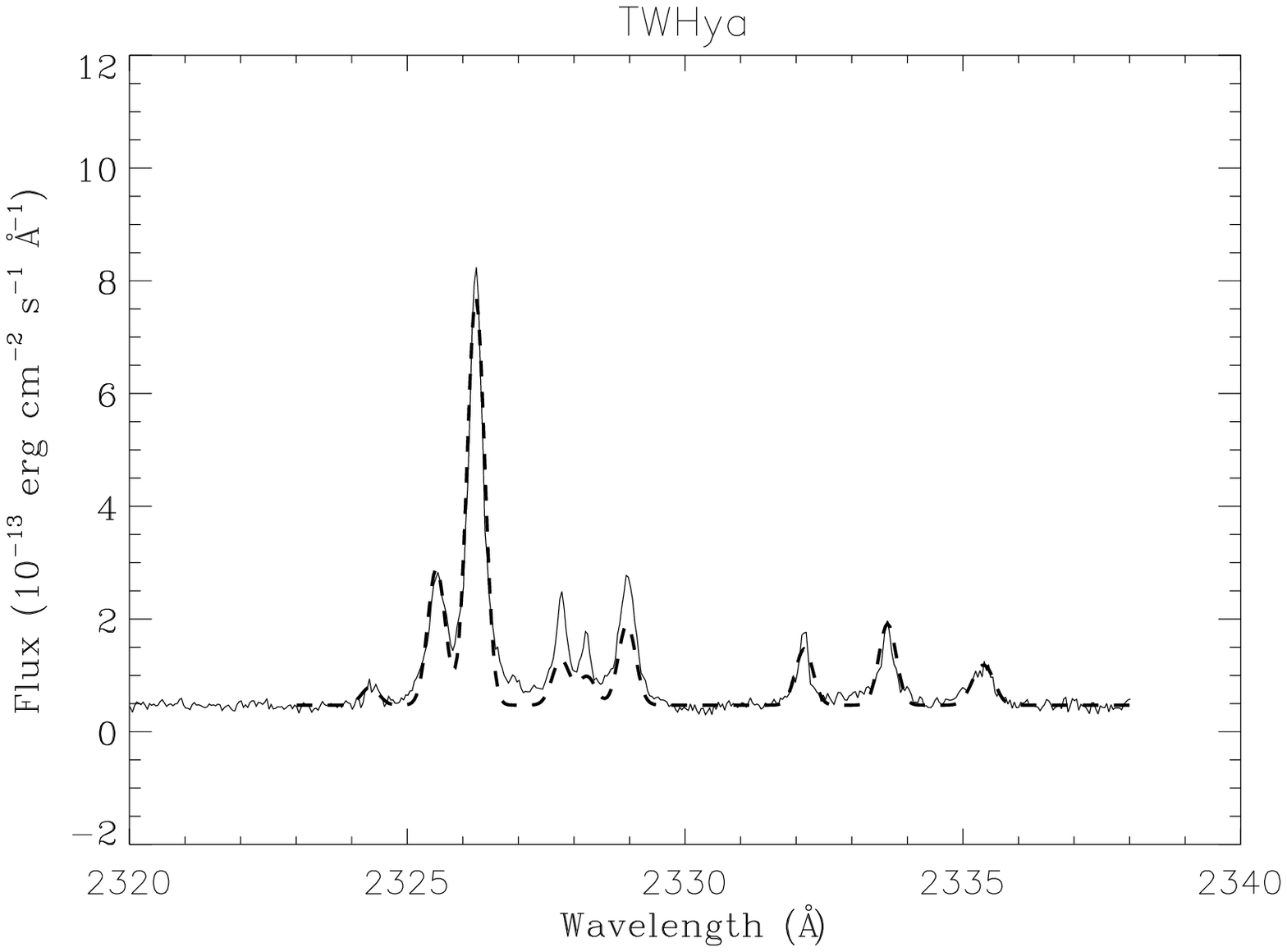}
\includegraphics[width=8cm]{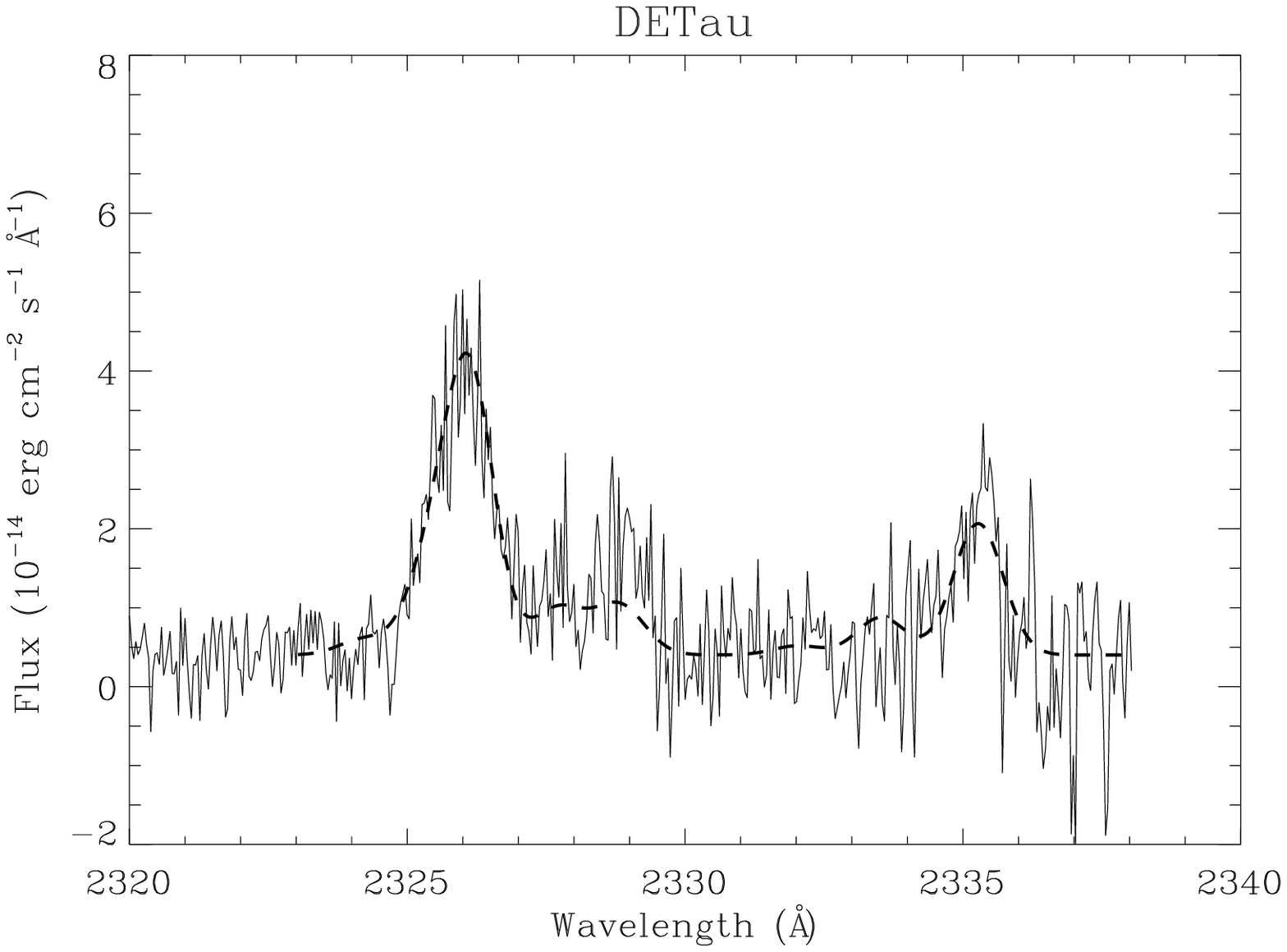}
\caption{Original spectra (solid lines) and their best fits (dotted lines) for two example stars: TW~Hya (top) and DE~Tau (bottom). \label{f6}}
\end{figure}

Fig.\ref{f6} shows two illustrative examples of the results of the fitting procedure. The targets selected are TW~Hya, with high S/N and 
DE~Tau with low S/N. The difference in S/N is readily  observed in the $\chi^2 $ surface (see Fig.\ref{chi}); the height of the surface above the ($n_{\rm e}, T_{\rm e}$) 
plane increases as the S/N decreases.  However both surfaces share some common characteristics: (1) a steep rise of the $\chi^2$ surface towards low 
$T_{\rm e}$ and low $n_{\rm e}$ and (2) there is always a narrow range of ($n_{\rm e},T_{\rm e}$)
that gives the best statistical fits to the original data (see the projected contours of the $\chi^2 $ surfaces on the $n_{\rm e}, T_{\rm e}$ plane in Fig.\ref{chi}). 

\subsection{($n_{\rm e},T_{\rm e}$) in the line emission region}

\begin{figure}
\includegraphics[width=8cm]{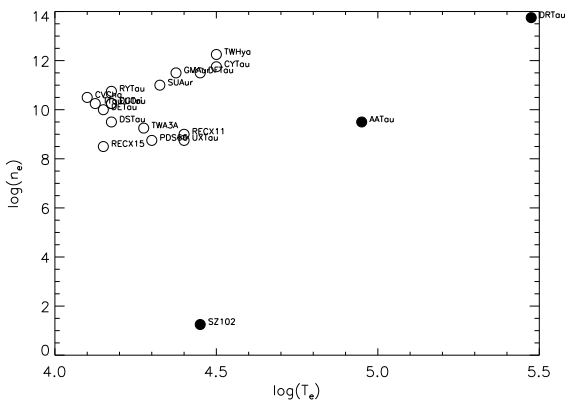}
\caption{Electron densities ($n_{\rm e}$ in cm$^{-3}$) and temperatures ($T_{\rm e}$ in K)
corresponding to the best fit to the observed spectra for
the stars in the sample. Circle radius corresponds to the uncertainties associated with $n_{\rm e}$ and $T_{\rm e}$.
Filled circles indicate stars with 
values out of the range where most of the sources in the sample are present. \label{tene}}
\end{figure}

Fig.\ref{tene} shows the electron densities and temperatures
corresponding to the optimal  fits. 
For stars with multiple observations, only the best-fitting (with the minimum $\chi_{opt}^2$) results are plotted. 
The differences among observations are small having very similar results in most of the
cases (see Table~\ref{tab3}).

Most  sources are grouped in a region with 
$4.1\la\log(T_{\rm e})\la 4.5$ and
$8\la\log(n_{\rm e})\la 12$.
There are three stars outside this region:
DR~Tau, AA~Tau and SZ~102. DR~Tau converged to values
lying very close to the limits of the $n_{\rm e}-T_{\rm e}$ grid. 
In the case of SZ~102, the low density probably
indicates that the C~II] emission is dominated by an extended ionized envelope.
Something similar might be occurring in AA~Tau,
a CTTS with a warped disc \citep{menard2003} that displays very peculiar
profiles in the UV emission lines \citep{france2012,ardila2013,aig2013b}. 
These three stars are 
represented in the figure as filled circles.
These ``unusual" values lead us to think that maybe the C~II], Fe~II] and Si~II]
lines are not formed under the same physical conditions as
the other sources. Therefore, these three stars are excluded
from the following analysis.

\subsection{Consistency tests}
\label{tests}

For this purpose, we have compared the observed flux
in the  C~II] feature with the flux derived from the best fitting model for each target - including the C~II] quintuplet
and the unresolved Fe~II]$_{2328.1}$ and Si~II]$_{2329.23}$
lines.

The observed flux has been measured in the
range 2324-2330~\AA\ as
$F_{obs} = (f-N_{pix} F_{cont}) \Delta\lambda$, where
$F_{cont}$ is the continuum average flux, $N_{pix}$ is
the number of pixels in the selected window (151
pixels), $f$ is the wavelength-integrated line flux
and $\Delta\lambda$ the step in wavelength (0.04~\AA).
We also estimated the corresponding flux error as
$\delta F = N_{pix} \cdot \Delta\lambda \cdot \sigma_{cont}$ (being $\sigma _{cont}$
the dispersion around this average).
The continuum was measured in the 2320-2323~\AA\ spectral
range. 

The simulated flux of each line has been calculated as the 
integral of the Gaussian function fitting that line.
Table~\ref{tab4} shows the fluxes for each line. 
The total flux of the C~II] quintuplet has been calculated from the
best-fitting models as
\begin{equation}
\label{eq4}
F_{sim}(C{\rm II}]) =  \sigma \sqrt{2 \pi} F_0 \displaystyle\sum_{i=0}^{4} R_i
\end{equation}
\noindent
The  Si~II]$_{2335}$ flux is the sum of the components at 2335.12 and 2335.52~\AA\
since they are not resolved in the \textit{HST}/STIS spectra.
\begin{table*}
\begin{flushleft}
\caption{Fluxes of the main features derived from the fitting procedure$^{(a)}$ \label{tab4}}
\begin{tabular}{ccccccc}
\hline
Star    & Data set     & Flux(C II])   &  Flux(Fe II]$_{2328}$)   &  Flux(Si II]$_{2329}$)    & Flux(Fe II]$_{2333}$) & Flux(Si II]$_{2335}$) \\
\cline{3-7}
& & \multicolumn{5}{c}{(${\rm erg}\ {\rm s}^{-1}\ {\rm cm}^{-2}$)} \\
\hline
AA Tau	&	ob6ba7030	&	($	9.02	\pm	0.17	)	\times 10^{-14}$	&	($	2.58	\pm	1.12	)	\times 10^{-20}$	&	($	4.59	\pm	1.42	)	\times 10^{-18}$	&	($	7.47	\pm	3.22	)	\times 10^{-20}$	&	($	4.45	\pm	1.42	)	\times 10^{-16}$	\\ \hline
CV Cha	&	ob6b18020	&	($	4.84	\pm	1.02	)	\times 10^{-14}$	&	($	5.09	\pm	1.50	)	\times 10^{-15}$	&	($	1.99	\pm	0.36	)	\times 10^{-16}$	&	($	1.46	\pm	0.43	)	\times 10^{-14}$	&	($	3.52	\pm	0.59	)	\times 10^{-14}$	\\ \hline
CY Tau	&	o5cf0302	&	($	4.73	\pm	0.03	)	\times 10^{-14}$	&	($	6.36	\pm	4.02	)	\times 10^{-16}$	&	($	1.52	\pm	0.22	)	\times 10^{-17}$	&	($	1.81	\pm	1.14	)	\times 10^{-15}$	&	($	3.16	\pm	0.46	)	\times 10^{-15}$	\\ \hline
DE Tau	&	ob6ba8030	&	($	6.39	\pm	0.29	)	\times 10^{-14}$	&	($	1.82	\pm	0.50	)	\times 10^{-15}$	&	($	1.30	\pm	0.26	)	\times 10^{-16}$	&	($	5.21	\pm	1.44	)	\times 10^{-15}$	&	($	1.86	\pm	0.38	)	\times 10^{-14}$	\\ \hline
DF Tau	&	o5kc0102	&	($	2.43	\pm	0.03	)	\times 10^{-13}$	&	($	3.75	\pm	2.34	)	\times 10^{-15}$	&	($	1.09	\pm	0.17	)	\times 10^{-16}$	&	($	1.07	\pm	0.67	)	\times 10^{-14}$	&	($	2.24	\pm	0.35	)	\times 10^{-14}$	\\ \hline
DG Tau	&	o63l03020	&	($	3.13	\pm	0.09	)	\times 10^{-14}$	&	($	8.78	\pm	2.52	)	\times 10^{-16}$	&	($	5.03	\pm	0.70	)	\times 10^{-17}$	&	($	2.51	\pm	0.72	)	\times 10^{-15}$	&	($	8.03	\pm	1.26	)	\times 10^{-15}$	\\ 
	&	o63l03030	&	($	1.48	\pm	0.13	)	\times 10^{-14}$	&	($	9.91	\pm	2.82	)	\times 10^{-16}$	&	($	7.07	\pm	1.44	)	\times 10^{-17}$	&	($	2.84	\pm	0.81	)	\times 10^{-15}$	&	($	1.01	\pm	0.21	)	\times 10^{-14}$	\\ \hline
DR Tau	&	o63l04010	&	($	4.53	\pm	0.01	)	\times 10^{-14}$	&	($	3.39	\pm	2.17	)	\times 10^{-19}$	&	($	2.18	\pm	0.13	)	\times 10^{-17}$	&	($	9.94	\pm	6.35	)	\times 10^{-19}$	&	($	4.63	\pm	0.27	)	\times 10^{-15}$	\\ 
	&	o63l04020	&	($	4.76	\pm	0.00	)	\times 10^{-14}$	&	($	3.56	\pm	2.28	)	\times 10^{-19}$	&	($	2.29	\pm	0.14	)	\times 10^{-17}$	&	($	1.05	\pm	0.67	)	\times 10^{-18}$	&	($	4.87	\pm	0.29	)	\times 10^{-15}$	\\ \hline
DS Tau	&	o5cf01020	&	($	7.43	\pm	0.06	)	\times 10^{-14}$	&	($	3.73	\pm	1.13	)	\times 10^{-16}$	&	($	6.24	\pm	0.79	)	\times 10^{-17}$	&	($	1.06	\pm	0.32	)	\times 10^{-15}$	&	($	7.58	\pm	0.98	)	\times 10^{-15}$	\\ 
	&	o63l08010	&	($	6.40	\pm	0.08	)	\times 10^{-14}$	&	($	5.91	\pm	1.06	)	\times 10^{-16}$	&	($	6.25	\pm	0.99	)	\times 10^{-17}$	&	($	1.69	\pm	0.31	)	\times 10^{-15}$	&	($	7.00	\pm	1.06	)	\times 10^{-15}$	\\ 
	&	o63l08020	&	($	3.74	\pm	0.08	)	\times 10^{-14}$	&	($	6.49	\pm	1.08	)	\times 10^{-16}$	&	($	6.87	\pm	1.17	)	\times 10^{-17}$	&	($	1.86	\pm	0.31	)	\times 10^{-15}$	&	($	7.47	\pm	1.13	)	\times 10^{-15}$	\\ \hline
FU Ori	&	o63l07020	&	($	2.77	\pm	0.06	)	\times 10^{-14}$	&	($	7.77	\pm	2.31	)	\times 10^{-16}$	&	($	4.45	\pm	0.66	)	\times 10^{-17}$	&	($	2.22	\pm	0.66	)	\times 10^{-15}$	&	($	7.11	\pm	1.18	)	\times 10^{-15}$	\\ \hline
GM Aur	&	ob6ba1030	&	($	7.32	\pm	0.29	)	\times 10^{-14}$	&	($	3.31	\pm	1.84	)	\times 10^{-15}$	&	($	5.56	\pm	0.63	)	\times 10^{-17}$	&	($	9.40	\pm	5.24	)	\times 10^{-15}$	&	($	1.14	\pm	0.13	)	\times 10^{-14}$	\\ \hline
PDS66	&	ob6b23030	&	($	2.85	\pm	0.06	)	\times 10^{-13}$	&	($	2.00	\pm	0.37	)	\times 10^{-15}$	&	($	3.59	\pm	0.33	)	\times 10^{-16}$	&	($	5.68	\pm	1.06	)	\times 10^{-15}$	&	($	3.15	\pm	0.27	)	\times 10^{-14}$	\\ \hline
RECX15	&	ob6bb7030	&	($	8.39	\pm	0.22	)	\times 10^{-14}$	&	($	1.60	\pm	0.37	)	\times 10^{-15}$	&	($	2.19	\pm	0.46	)	\times 10^{-16}$	&	($	4.57	\pm	1.07	)	\times 10^{-15}$	&	($	1.88	\pm	0.39	)	\times 10^{-14}$	\\ \hline
RECX11	&	ob6bc4030	&	($	9.06	\pm	0.13	)	\times 10^{-14}$	&	($	1.87	\pm	0.59	)	\times 10^{-16}$	&	($	6.39	\pm	0.97	)	\times 10^{-17}$	&	($	5.26	\pm	1.70	)	\times 10^{-16}$	&	($	5.83	\pm	0.88	)	\times 10^{-15}$	\\ \hline
RY Tau	&	o63l01010	&	($	9.20	\pm	0.50	)	\times 10^{-14}$	&	($	2.43	\pm	0.69	)	\times 10^{-15}$	&	($	3.25	\pm	0.84	)	\times 10^{-16}$	&	($	6.96	\pm	2.00	)	\times 10^{-15}$	&	($	2.78	\pm	0.72	)	\times 10^{-14}$	\\ 
	&	o63l01020	&	($	8.65	\pm	0.52	)	\times 10^{-14}$	&	($	4.84	\pm	1.79	)	\times 10^{-15}$	&	($	1.46	\pm	0.18	)	\times 10^{-16}$	&	($	1.38	\pm	0.51	)	\times 10^{-14}$	&	($	2.74	\pm	0.36	)	\times 10^{-14}$	\\ 
	&	o63l01030	&	($	8.58	\pm	0.45	)	\times 10^{-14}$	&	($	5.48	\pm	2.16	)	\times 10^{-15}$	&	($	1.17	\pm	0.08	)	\times 10^{-16}$	&	($	1.56	\pm	0.62	)	\times 10^{-14}$	&	($	2.30	\pm	0.16	)	\times 10^{-14}$	\\ \hline
SU Aur	&	o63l05010	&	($	6.70	\pm	0.13	)	\times 10^{-14}$	&	($	2.05	\pm	1.03	)	\times 10^{-15}$	&	($	6.29	\pm	0.59	)	\times 10^{-17}$	&	($	5.84	\pm	2.92	)	\times 10^{-15}$	&	($	1.22	\pm	0.13	)	\times 10^{-14}$	\\ 
	&	o63l05020	&	($	6.96	\pm	0.13	)	\times 10^{-14}$	&	($	1.96	\pm	0.59	)	\times 10^{-15}$	&	($	1.12	\pm	0.17	)	\times 10^{-16}$	&	($	5.59	\pm	1.70	)	\times 10^{-15}$	&	($	1.79	\pm	0.31	)	\times 10^{-14}$	\\ 
	&	ob6bb1030	&	($	6.09	\pm	0.26	)	\times 10^{-14}$	&	($	2.32	\pm	0.78	)	\times 10^{-15}$	&	($	9.91	\pm	1.32	)	\times 10^{-17}$	&	($	6.64	\pm	2.22	)	\times 10^{-15}$	&	($	1.73	\pm	0.26	)	\times 10^{-14}$	\\ \hline
SZ102	&	ob6bb9030	&	($	5.97	\pm	0.08	)	\times 10^{-14}$	&	($	2.02	\pm	0.67	)	\times 10^{-18}$	&	($	3.67	\pm	0.60	)	\times 10^{-17}$	&	($	1.44	\pm	0.47	)	\times 10^{-17}$	&	($	1.94	\pm	0.32	)	\times 10^{-15}$	\\ \hline
T Tau	&	o63l02010	&	($	1.83	\pm	0.13	)	\times 10^{-13}$	&	($	8.89	\pm	3.09	)	\times 10^{-15}$	&	($	3.70	\pm	0.64	)	\times 10^{-16}$	&	($	2.55	\pm	0.89	)	\times 10^{-14}$	&	($	6.49	\pm	1.18	)	\times 10^{-14}$	\\ 
	&	o63l02020	&	($	1.78	\pm	0.20	)	\times 10^{-13}$	&	($	8.89	\pm	2.72	)	\times 10^{-15}$	&	($	4.81	\pm	0.99	)	\times 10^{-16}$	&	($	2.55	\pm	0.78	)	\times 10^{-14}$	&	($	7.75	\pm	1.58	)	\times 10^{-14}$	\\ 
	&	o63l02030	&	($	1.77	\pm	0.20	)	\times 10^{-13}$	&	($	8.83	\pm	2.70	)	\times 10^{-15}$	&	($	4.78	\pm	0.98	)	\times 10^{-16}$	&	($	2.53	\pm	0.78	)	\times 10^{-14}$	&	($	7.69	\pm	1.56	)	\times 10^{-14}$	\\ \hline
TWA3A	&	ob6b22030	&	($	1.18	\pm	0.03	)	\times 10^{-13}$	&	($	1.07	\pm	0.16	)	\times 10^{-15}$	&	($	1.53	\pm	0.14	)	\times 10^{-16}$	&	($	3.04	\pm	0.45	)	\times 10^{-15}$	&	($	1.52	\pm	0.11	)	\times 10^{-14}$	\\ \hline
TW Hya	&	o59d01020	&	($	4.80	\pm	0.19	)	\times 10^{-13}$	&	($	1.99	\pm	1.14	)	\times 10^{-14}$	&	($	1.61	\pm	0.20	)	\times 10^{-16}$	&	($	5.65	\pm	3.25	)	\times 10^{-14}$	&	($	3.39	\pm	0.43	)	\times 10^{-14}$	\\ \hline
UX Tau	&	ob6b54030	&	($	3.94	\pm	0.07	)	\times 10^{-14}$	&	($	7.86	\pm	2.54	)	\times 10^{-17}$	&	($	2.83	\pm	0.43	)	\times 10^{-17}$	&	($	2.23	\pm	0.72	)	\times 10^{-16}$	&	($	2.47	\pm	0.37	)	\times 10^{-15}$	\\ \hline

\end{tabular}
\begin{tabular}{ll}
$^{(a)}$ & Fluxes are not extinction corrected. \\
\end{tabular}
\end{flushleft}
\end{table*}

\begin{figure}
\includegraphics[width=8cm]{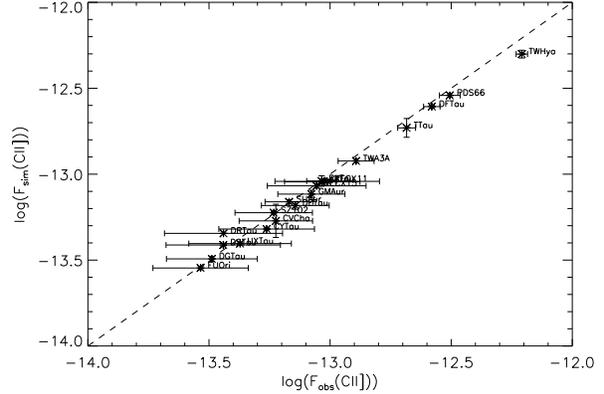}
\caption{The observed flux in the 2326~\AA\ feature 
compared with the derived from the best fit. Dashed line marks
the 1:1 relation. \label{obssimsnr}}

\end{figure}

The comparison between observed and fitted flux is shown in Fig.\ref{obssimsnr}.
Most of the observed 
fluxes are slightly higher than the simulated ones but the discrepancy is well within the
expected value given the S/N of the data. TW~Hya shows the largest discrepancy
that we interpret as a result of the simplicity of the modelling, i. e. the difficulties to fit the 
data to a ``single plasma" emission. In this sense, we would like to remark that the $(n_{\rm e}, T_{\rm e})$ 
values in Table~\ref{tab3} should be understood as average values on the plasma emission region.

We have also calculated the contribution of the Fe~II]$_{2328}$ and Si~II]$_{2329}$ fluxes
to the 2326~\AA\ feature, unresolved in most of the TTSs spectra. From the simulated
spectra, we have found that  Fe~II]$_{2328}$ emission can account for up to $\sim 15$ per cent 
of the flux, whereas Si~II]$_{2329}$ contribution is negligible  ($\la 0.5$ per cent ).

\subsubsection{Line ratios as $T_{\rm e}$ and $n_{\rm e}$ indicators}
\label{secratios}

\begin{figure}
\includegraphics[width=8cm]{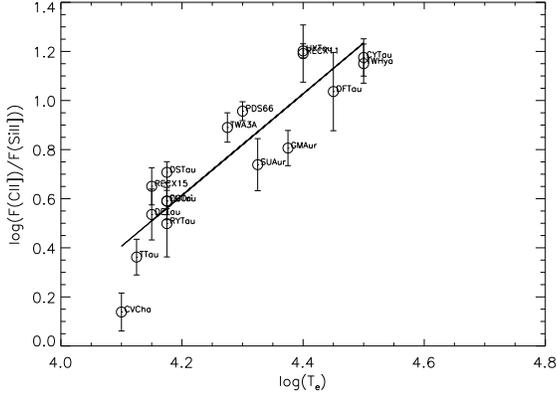}
\caption{The ratio between C~II] and Si~II] fluxes
$F({\rm CII}])/F({\rm SiII}])$ as a function of the
temperature $T_{\rm e}$ (K). Solid line is the best linear
fit. \label{ciisiite}}
\end{figure}

The C~II]/Si~II]  flux ratio is  a sensitive tracer of the
electron temperature in the range of interest. As it is shown in Fig.\ref{ciisiite},
$T_{\rm e}$ is basically derived from this ratio in our code.
The regression line in Fig.\ref{ciisiite} has a Pearson's coefficient of $r=0.91$ with
a $p{\rm -value}$\footnote{
$p{\rm -value}=p$ means that, for a random population there is $100 \cdot p$ per cent
probability that the cross-correlation
coefficient will be $r$ or better.
We are assuming that the correlation coefficient is
statistically significant if the $p{\rm -value}$ is lower than 5 per cent.}
$=4.8 \times 10^{-7}$.
The regression equation is:
\begin{equation}
\log(F({\rm CII]})/F({\rm SiII]})) = (2.1 \pm 0.3) \, \log(T_{\rm e}) - (8.1 \pm 1.1)
\end{equation}

We have not found any significant correlation
between C~II]/Fe~II]$_{2333}$ flux ratio and the temperature.

\begin{figure}
\includegraphics[width=8cm]{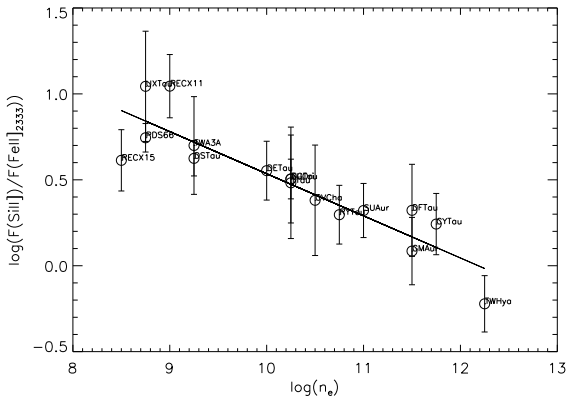}
\caption{
$F({\rm SiII}])/F({\rm FeII}])$ as a function of the electron
density $n_{\rm e}$ (cm$^{-3}$). Solid line represents the best linear fit.\label{siifeiine}}
\centering

\end{figure}

Regarding electron density, we have recovered the expected relation between $n_{\rm e}$ and the 
C~II]/Fe~II]$_{2333}$ and Si~II]/Fe~II]$_{2333}$ ratios. 
The regression parameters are:
\begin{itemize}
\item For C~II]/Fe~II]$_{2333}$: $r=-0.6$ and $p{\rm -value}=0.015$
\item For Si~II]/Fe~II]$_{2333}$: $r=-0.9$, $p{\rm -value}=8.34 \times 10^{-7}$ and regression equation:
$\log(F({\rm SiII]})/F({\rm FeII]})) =  (-0.25 \pm 0.03) \, \log(n_{\rm e}) + (3.02 \pm 0.32)$, as shown in Fig.\ref{siifeiine}.
\end{itemize}

\section{{\rm C~II]} as an accretion tracer}
\label{ciimdot}

The C~II] quintuplet have been found to be a good tracer of the accretion rate \citep{calvet2004,ingleby2013}.
In this section, we discuss this point as well as the relationship between the obtained results, 
($n_{\rm e}$, $T_{\rm e}$, $\sigma$) and accretion rate ($\dot{M}$).

\subsection{Dispersion versus electron temperature}

\begin{figure}
\includegraphics[width=8cm]{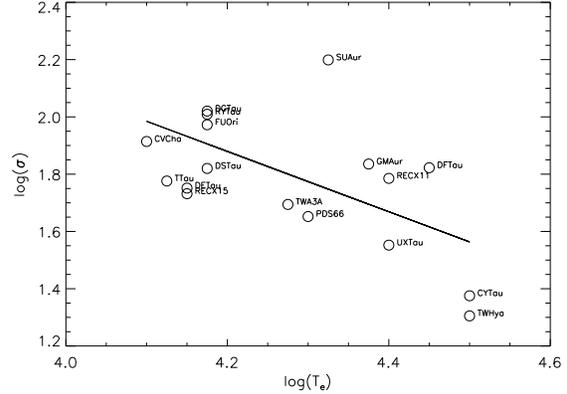}
\caption{Line dispersion $\sigma$~(km~s$^{-1}$) as a function of
temperature $T_{\rm e}$~(K). Solid line is the
best linear fit. The error bars for $\log (\sigma)$ are smaller than the circle size. \label{sigmatene}} 

\end{figure}

Further insight on the source of the profile broadening can be drawn from 
Fig.\ref{sigmatene}. 
The line dispersions that best fit the observed spectra are shown in Table~\ref{tab3} and 
they are in the range $20 \la \sigma \la 160$~km~s$^{-1}$. 
TW~Hya and CY~Tau have $\sigma < 25$~km~s$^{-1}$ and high $T_{\rm e}$ values
($\log T_{\rm e}({\rm K}) \simeq 4.4-4.5$). For these stars the line broadening is
consistent with thermal broadening ($v_{th} \sim 22$~km~s$^{-1}$). 
SU~Aur is the source with the largest line broadening, $\sigma > 100$~km~s$^{-1}$, 
and a temperature of $T_{\rm e} \simeq 10^{4.3}$~K. This star is the fastest rotator 
in the sample ($v \, \sin i \sim 60$~km~s$^{-1}$) thus, rotation could be
an important source of line broadening.
The rest of the stars have intermediate $\sigma$ values ($40 \la \sigma \la 100$~km~s$^{-1}$) and 
temperatures in the range $\log T_{\rm e}({\rm K})\simeq 4.1-4.45$. 
The dispersions are suprathermal and the contribution of
rotational broadening is negligible since with $v \, \sin i$ values are in the range 
$\sim 5-25$~km~s$^{-1}$ (see Table~\ref{biblio}).
There is a mild correlation between $\sigma$ and $T_{\rm e}$, 
as shown in Fig.\ref{sigmatene}
($r=-0.6$ and a $p{\rm -value}=0.018$).

\subsection{Dispersion versus accretion rate}

\begin{figure}
\includegraphics[width=8cm]{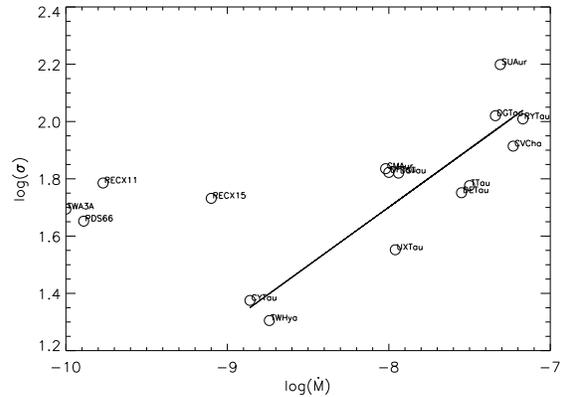}
\caption{The calculated line width $\sigma$~(km~s$^{-1}$) as a
function of the stellar accretion rate $\dot{M}$~(M$_{\sun}$~yr$^{-1}$)
(taken from the literature). Solid line is the
best linear fit for stars with
$\dot{M} \geq 10^9$~M$_{\sun}$~yr$^{-1}$. The error bars for $\log (\sigma)$ are smaller than the circle size. \label{sigmamacr}}

\end{figure}

We have also examined the relation between dispersion, $\sigma$ and 
accretion rate, $\dot{M}$. As shown in Fig.\ref{sigmamacr}, TTSs show a statistically significant
correlation between $\sigma$ and $\dot{M}$: the higher
the accretion rate the wider the line.
Note that there is a small group
of TTSs (TWA~3A, RECX~11, RECX~15 and PDS~66) with $\dot{M} < 10^{-9}$~M$_{\sun}$~yr$^{-1}$,
that seem to have too low accretion rates
for the given dispersion. PDS~66 also displays an unusually high C~II] flux for the
accretion rate derived by \citet{ingleby2013}. For this reason 
these stars have not been considered to determine the correlation coefficient.
The Pearson's coefficient is $r=0.87$ with a
$p{\rm -value}=0.0002$.
This trend suggests a clear connection between the region in which lines are formed and the accretion 
process and agrees with those trends reported recently for other UV spectral tracers
\citep{ardila2013,aig2013b}. 

\subsection{C~II] luminosity versus accretion rate}

Here we re-examine the correlation reported by \citet{ingleby2013} from low-dispersion data
between the accretion rate/luminosity  and the C~II] flux.
Fluxes are extinction corrected according
to \citet{valencic2004} assuming $R_V=3.1$
(see Table~\ref{biblio} for a compilation of the 
$A_V$ values and distances used in the calculation,
as well as other relevant parameters).
The extinction $A_V$ is one of the
major sources of uncertainty affecting, among 
other things, the accretion rate  estimates. 
For this reason, extinctions have been selected
mainly from the same source than the accretion rates
\citep{ingleby2013}. As a test, we have repeated
the analysis with data from \citet{ardila2013},
and found the same general trend.
\begin{figure}
\includegraphics[width=8cm]{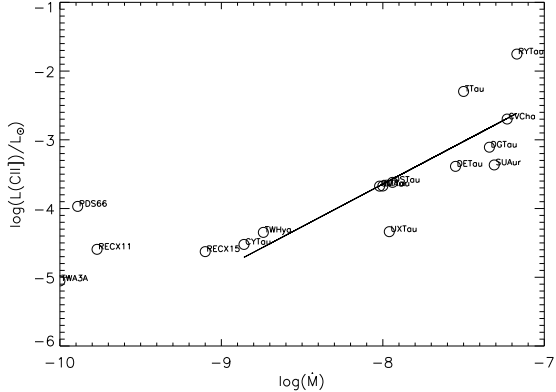}
\caption{The C~II] luminosity (in L$_{\sun}$) as a
function of the accretion rate (M$_{\sun}$~yr$^{-1}$). Solid line is the
best linear fit for stars with
$\dot{M} \geq 10^9$~M$_{\sun}$~yr$^{-1}$. \label{ciiavmacr}}
\end{figure}
As shown in Fig.\ref{ciiavmacr}, the C~II] luminosity increases as the accretion rate does: 
\begin{equation}
\log(L({\rm CII]})/{\rm L}_{\sun})=(1.24 \pm 0.26) \log{\dot{M}} + (6.27 \pm 2.06)
\end{equation}
\noindent
with a Pearson's correlation coefficient of $r=0.83$ ($p{\rm -value}=0.0008$).
This correlation is for stars with $\dot{M} > 10^{-9}$~M$_{\sun}$~yr$^{-1}$.
For comparison, \citet{ingleby2013} obtain a slope $ \simeq 0.9 \pm 0.2$
from low-dispersion data.

\subsection{Electron density versus accretion rate}

\begin{figure}
\includegraphics[width=8cm]{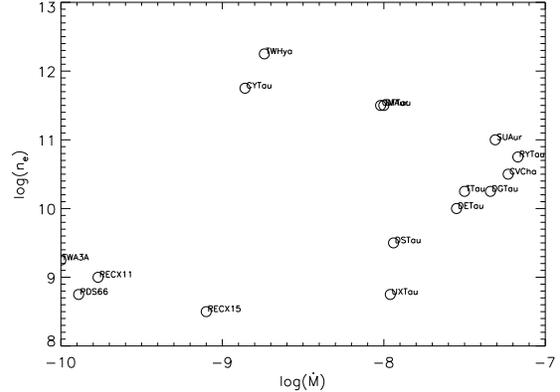}
\caption{Electron density $n_{\rm e}$ (cm$^{-3}$) as a function of the
accretion rate $\dot{M}$~(M$_{\sun}$~yr$^{-1}$). \label{nemacr}}

\end{figure}

In Fig.\ref{nemacr} we have plotted the electron  density as a function of the accretion rate.
TWA~3A, RECX~11, RECX~15 and PDS~66 have again a peculiar behaviour related
with their,  apparently, too low accretion rates when compared with the observed electron density  
in the emission region.
There are four stars (TW~Hya, CY~Tau, GM~Aur and DF~Tau)
with $n_{\rm e} > 10^{11}$~cm$^{-3}$. There seems to be a trend for $n_{\rm e}$ to increase as the accretion rate does it 
($r=0.92$ and a $p{\rm -value}=0.001$) in sources with 
$n_{\rm e} \la 10^{11}$~cm$^{-3}$ and $\dot{M} > 10^{-8}$~(M$_{\sun}$~yr$^{-1}$). 

\begin{table*}
\caption{Properties of the sample taken from the literature.}
\label{biblio}
\centering
\begin{tabular}{ccccccccccc}
\hline
Star & SpT & $L$ & $R$ & $M$ & d & $\log(\dot{M})$ & $v\,\sin i$ & $A_V$ & $v_{rad}$ & Ref.\\
 &  & (L$_{\sun})$ & (R$_{\sun})$ & (M$_{\sun})$ & (pc) & (M$_{\sun} \,{\rm yr}^{-1}$) & (km~s$^{-1}$)& (mag) & (km s$^{-1}$) & \\
\hline
AA Tau	&	K7	&	1	&	2.1	&	0.8	&	140	&	-7.82	&	11	&	1.9	&	16.1	&	1,17	\\
CY Tau	&	M2	&	0.31	&	1.63	&	0.55	&	140	&	-8.86	&	10.6	&	0.03	&	19.1	&	2,5,3,18	\\ 
CV Cha	&	G9	&	3.1	&	2	&	1.5	&	160	&	-7.23	&	32	&	1.5	&	16.1	&	1,3,10	\\
DE Tau	&	M2	&	0.8	&	2.4	&	0.4	&	140	&	-7.55	&	10	&	0.9	&	14.9	&	1,9,17	\\
DF TauA	&	M1	&	0.56	&	3.37	&	0.68	&	140	&	-8	&	16.1	&	0.15	&	11	&	2,7,3,9	\\ 
DG Tau	&	K6	&	1.15	&	1	&	0.88	&	140	&	-7.34	&	20	&	1.41	&	15.4	&	2,7,10,18	\\ 
DR Tau	&	K5	&	0.4	&	1.1	&	0.9	&	140	&	-7.28	&	10	&	1.4	&	27.6	&	1,3,10	\\ 
DS Tau	&	K5	&	0.68	&	1.36	&	1.04	&	140	&	-7.94	&	10	&	0.9	&	16.3	&	2,7,17	\\ 
FU Ori	&	G0	&	---	&	---	&	---	&	450	&	---	&	---	&	--	&	28	&	15,19	\\ 
GM Aur	&	K7	&	1.2	&	2.3	&	0.8	&	140	&	-8.02	&	12.4	&	0.6	&	15	&	1,9,17	\\ 
PDS66	&	K1	&	0.9	&	1.3	&	1.1	&	86	&	-9.89	&	14	&	0.2	&	11.6	&	1,3,14	\\
RECX15	&	M3	&	0.1	&	0.9	&	0.3	&	97	&	-9.1	&	15.9	&	0	&	15.9	&	1,3,13	\\ 
RECX11	&	K5	&	0.6	&	1.4	&	1	&	97	&	-9.77	&	16.4	&	0	&	18	&	1,3,12	\\
RY Tau	&	G1	&	9.6	&	2.9	&	2	&	140	&	-7.17	&	48.7	&	2.2	&	16.5	&	7,17	\\ 
SU Aur	&	G1	&	7.8	&	2.6	&	1.7	&	140	&	-7.31	&	59	&	0.9	&	16	&	7,17,18	\\ 
SZ102	&	K0	&	---	&	---	&	0.75	&	200	&	-8.1	&	---	&	0.32	&	5	&	3,4	\\ 
T Tau	&	K0	&	7.29	&	2.9	&	2.11	&	140	&	-7.5	&	20.1	&	1.46	&	19.1	&	2,8,17	\\ 
TW Hya	&	K7	&	0.3	&	1.1	&	0.8	&	56	&	-8.74	&	5.8	&	0	&	13.5	&	1,3,16	\\ 
TWA3A	&	M3	&	0.4	&	1.8	&	0.3	&	50	&	-10	&	12	&	0	&	---	&	1,14	\\
UX TauA	&	K5	&	0.91	&	2.05	&	1.09	&	140	&	-7.96	&	25.4	&	0.26	&	15.6	&	2,3,6,11,17	\\ \hline
\end{tabular}
\begin{tabular}{l}
(1) \citet{ingleby2013}; (2) \citet{white2001}; (3) \citet{ardila2013}; (4) \citet{france2012}; (5) \citet{gullbring1998} \\
(6) \citet{andrews2011}; (7) \citet{salyk2013}; (8) \citet{calvet2004}; (9) \citet{clarke2000}; (10) \citet{johnskrull2000};\\
(11) \citet{preibisch1997}; (12)\citet{jayawardhana2006};(13) \citet{woitke2011}; (14) \citet{dasilva2009};\\ 
(15) \citet{petrov2008}; (16) \citet{herczeg2006}; (17) \citet{hartmann1986}; (18) \citet{nguyen2012};\\
(19) \citet{malaroda2006}.
\end{tabular}
\end{table*}
\subsection{Blueshifted profiles}

The shift of the lines, $\delta$, obtained from the fitting, was corrected to the stellar rest frame and 
it is provided in Table~\ref{tab3}; the radial velocities of the TTSs are compiled in Table~\ref{biblio}. 
Note that the pointing errors in the STIS data result 
in a  velocity uncertainty of 3~km~s$^{-1}$, negligible for the purpose of this work.
Most TTSs satisfy $-20 \la \delta \la 20$~km~s$^{-1}$;
however, there are three stars namely, DG~Tau, FU~Ori and RY~Tau with clearly blueshifted emission
at velocities of  -81.5, -73.5 and -47.1~km~s$^{-1}$, respectively. This blueshift indicates 
a contribution from the unresolved base of the jet. 

\section{Conclusions}
\label{results}

In this work, we have studied the semiforbidden lines of C~II],
Si~II] and Fe~II] in the 2310-2340~\AA\ spectral range for a
sample of 20 TTSs using 30 medium resolution spectra
obtained with \textit{HST}/STIS instrument. 

As the lines are blended in a broad feature in most sources, 
we have developed a numerical method to determine the 
properties of the line emission region assuming that the  
radiating plasma can be characterized by a 
single $T_{\rm e}$ and $n_{\rm e}$ pair, considering solar abundances.
This is the first work where $n_{\rm e}$ and $T_{\rm e}$ has been determined
for such a large sample of TTSs; previous works dealt with much
smaller samples \citep{aig2001,aig2003}.

In magnetospheric accretion, matter flows from the inner border of the circumstellar disc on the magnetospheric
surface to finally fall on to the star. Near the stellar surface a dense and hot shock is formed producing hot spots.
The sheared magnetosphere-disc boundary layer is expected to be very prone to the
development of turbulent flows. 

Within this overall picture there are four issues worth remarking.

\begin{itemize}

\item In most TTSs, the C~II], Si~II] and Fe~II] radiation seems to be produced in an extended magnetospheric structure
characterized by  $10^{8} \la n_{\rm e} \la 10^{12}$~cm$^{-3}$ and $10^{4.1} \la T_{\rm e} \la 10^{4.5}$~K. 
The line broadening is suprathermal except for two stars (TW~Hya and CY~Tau).
The dispersion depends on the electron temperature of the radiating plasma and on the
accretion rate, suggesting a connection 
between the line formation region and the accretion process.
This is consistent with the line radiation being dominated by the magnetospheric accretion flow, close to the disc.
For TW~Hya and CY~Tau, the densities and temperatures are higher than
for the rest of the stars and similar to the observed in atmospheres of cool stars \citep{brown1984,brooks2001}.
Also, the line broadening is thermal. 
Therefore, the observed emission lines in TW~Hya and CY~Tau are formed in a different region in the magnetospheric accretion flow (likely close to the star).
In good agreement with this picture, the density and temperature in the line formation region are below the theoretical
predictions for the density and temperature in the accretion shock ($n_{\rm e} \simeq 10^{13}$~cm$^{-3}$ and $T_{\rm e} \simeq 10^6$~K) 
and about the densities and temperatures
expected in the funnel flow \citep[$n_{\rm e} \simeq 10^9-10^{12}$~cm$^{-3}$ and $T_{\rm e} \simeq 5 \times 10^3-10^{4.5}$~K; see for example][]{calvet1998,muzerolle2001}.

\item There are three sources,  DG~Tau, FU~Ori and RY~Tau with blueshifted lines centroid. 
DG~Tau and RY~Tau have resolved jets and FU~Ori has a strong wind.
The large blueshifted velocities in these stars can be due to the contribution 
of the outflows to the C~II] lines, suggesting that the properties in the base of the outflow are 
similar to those in the base of the accretion stream.
The electron densities of the jet sources derived from the C~II],
Si~II] and Fe~II] lines agree well with previous estimates of electron densities at the base of the jet
\citep{aig2001,aig2003,aig2007}. The observations
agree with the predictions of hot disc winds \citep{aig2005}.
From the theoretical point of view, it is expected that both, the base of the jet and the foot-point of the 
accretion flow, share similar physical conditions \citep[see e.g.][]{mohanty2008}.  

\item The C~II] quintuplet can be used as a reliable tracer 
of the mass accretion rate on the star. C~II] luminosity increases
as the accretion rate does it in agreement with previous
results by \citet{calvet2004,ingleby2013}.
\end{itemize}

\section*{Acknowledgements}

The authors acknowledge support from the Spanish Ministry of Economy and Competitiveness through grant AYA2011-29754-C03-01. 
We also wish to thank an anonymous referee for her/his useful comments.


\appendix
\section{Variability of the C~II] profiles}

\begin{figure*}
\begin{tabular}{cc}
\includegraphics[width=8cm]{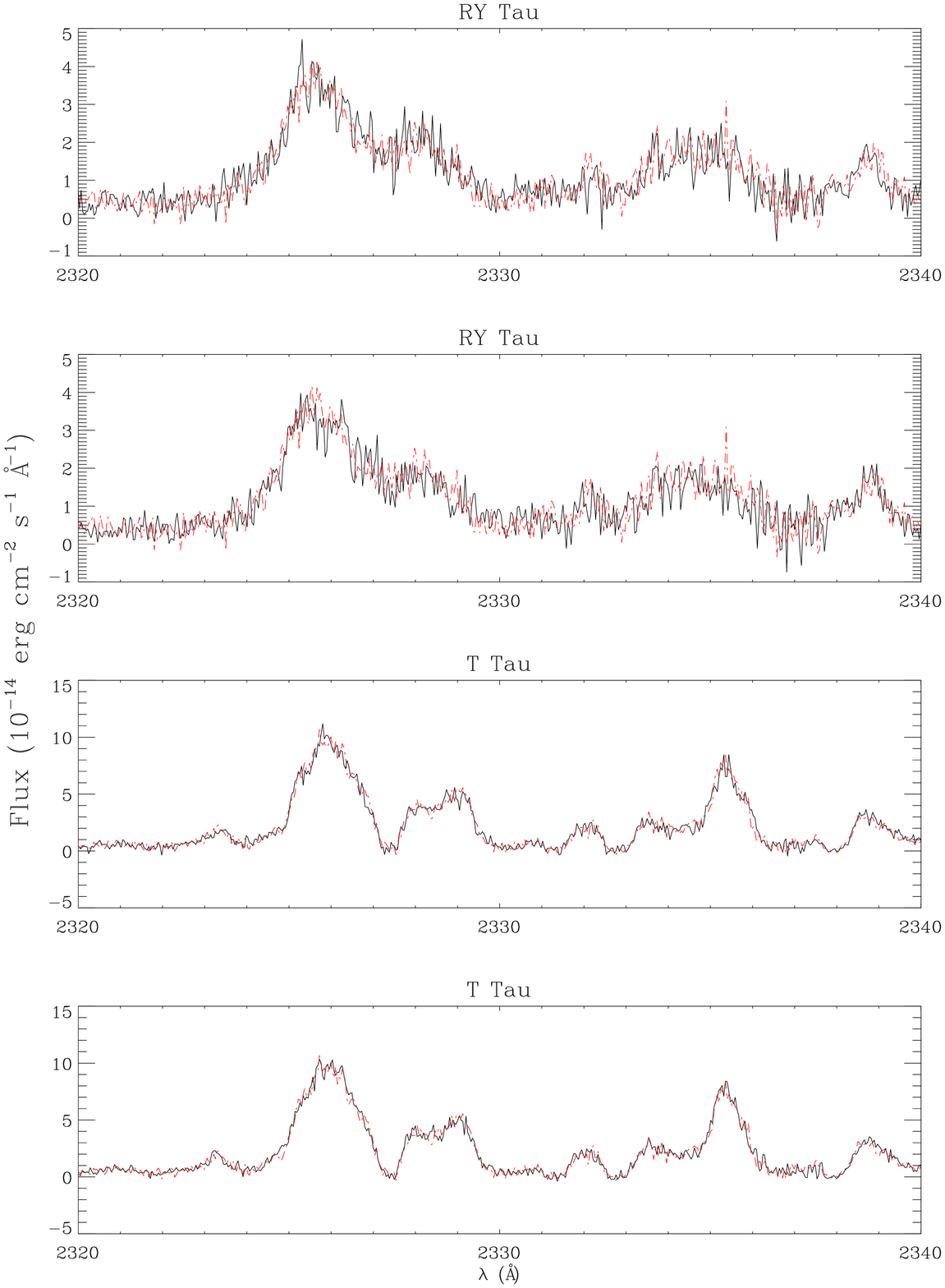} & \includegraphics[width=8cm]{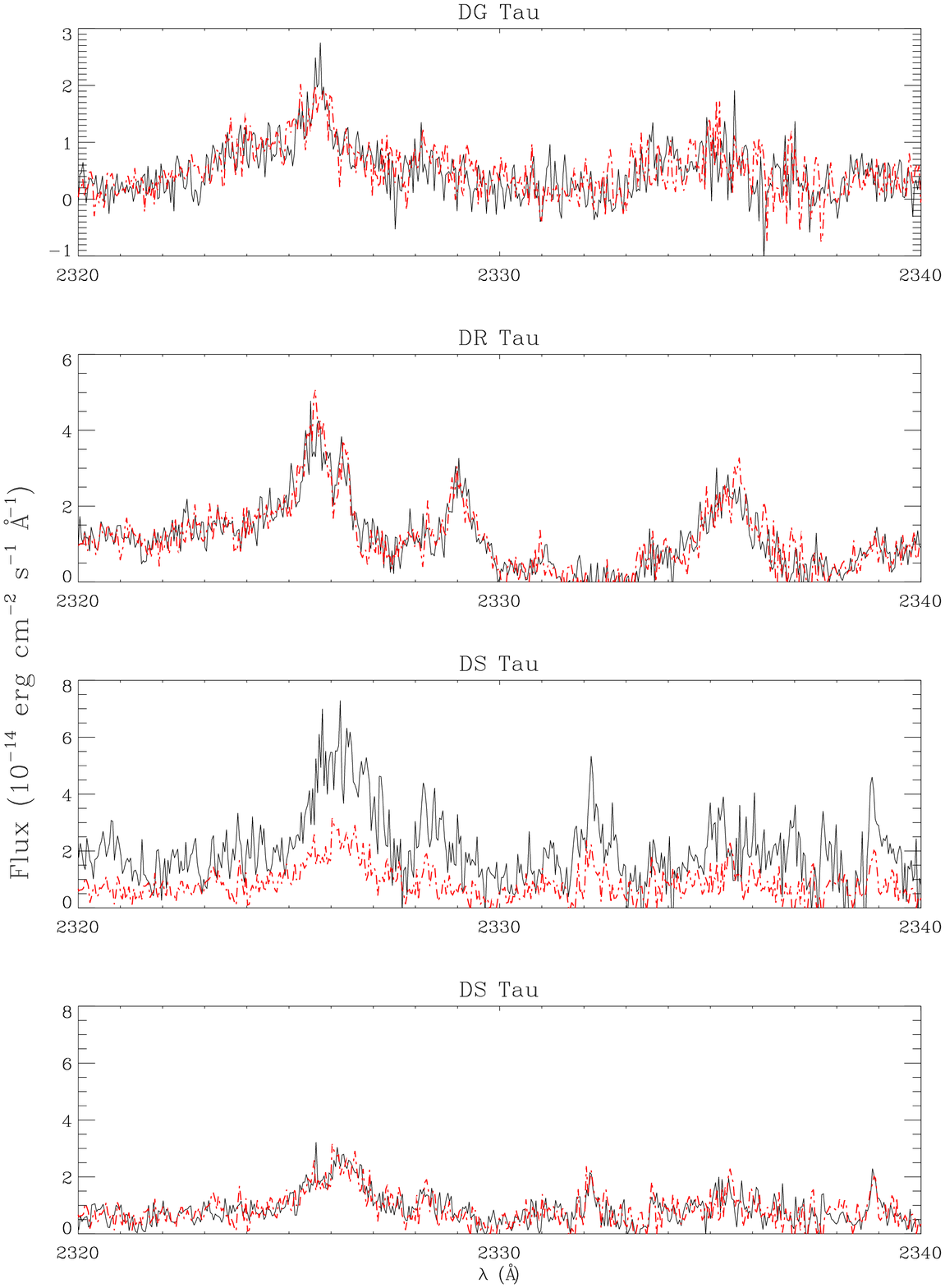} \\
\includegraphics[width=8cm]{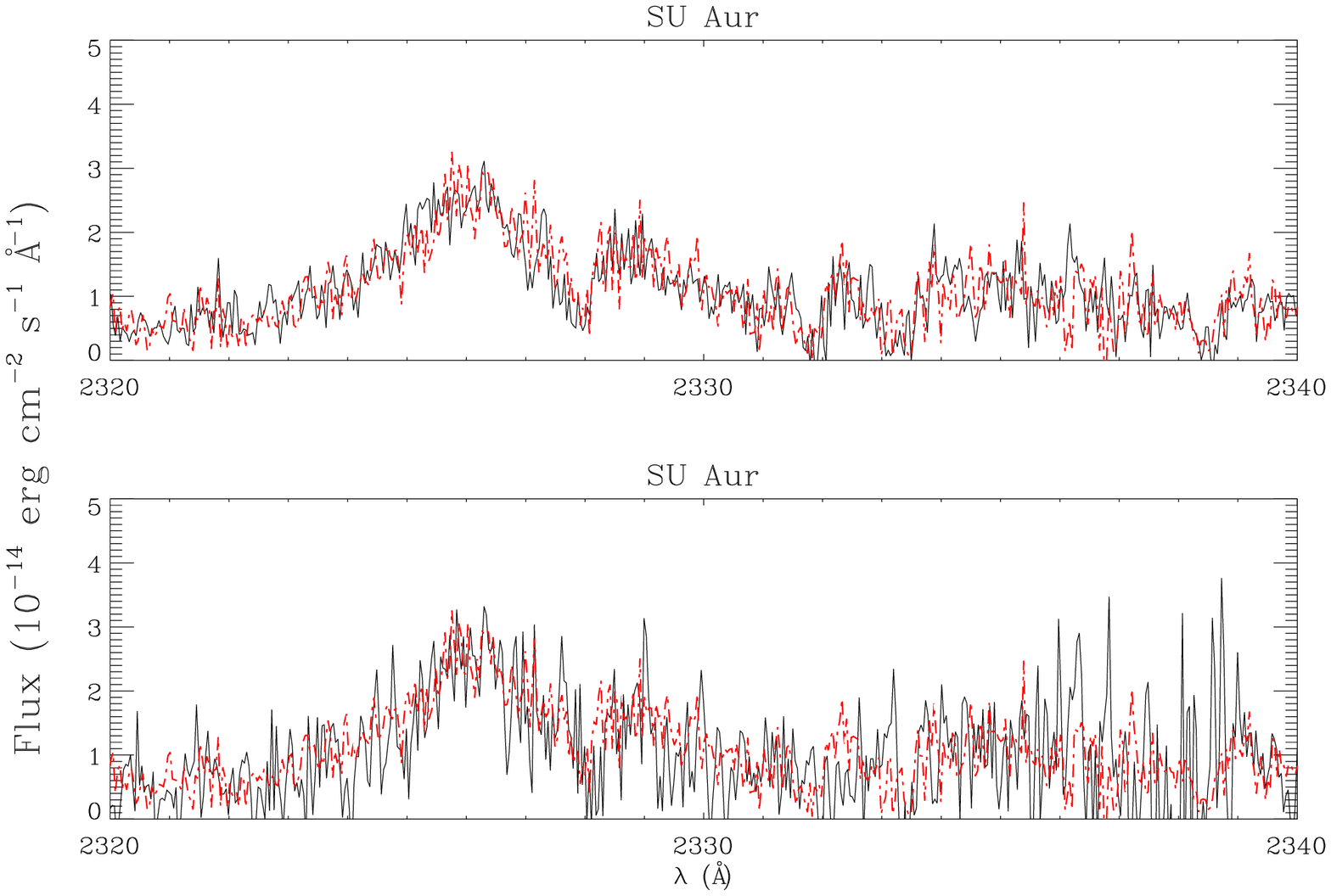} \\
\end{tabular}
\caption{Variability of the C~II profiles of the TTSs. For each star, the highest S/N observation is used as reference
and is superimposed as red dotted line. }
\label{varprof1}
\end{figure*}

Significant variations in the C~II] profiles are only found in DS~Tau (see Fig.\ref{varprof1}). 
In this section, we include the figures showing the variability of the C~II] profiles in TTSs. Only observations with S/N$>2$ are compared.

\end{document}